\documentclass[prb,twocolumn]{revtex4} 

\usepackage{graphicx}
\usepackage{amsmath}
\usepackage{amsfonts} 
\usepackage{mathrsfs} 
\usepackage{color}
\usepackage{natbib,hyperref,url}

\newcommand{\mn}{\mathbf}
\def\<{\langle}
\def\>{\rangle}

\def\beq{\begin{equation}}
\def\eeq{\end{equation}}
\def\barray{\begin{eqnarray}}
\def\earray{\end{eqnarray}}

\begin{document}

\title{Critical behavior of the XY model in complex topologies}
\author{Miguel Ib\'a\~nez Berganza  and Luca Leuzzi}
 \affiliation{IPCF-CNR, UOS Roma {\em Kerberos}. \\ Dipartimento di Fisica, Universit\`a ``La Sapienza''. \\ Piazzale A. Moro, 5, 00185 Roma.}
\email{miguel.berganza@roma1.infn.it}
\email{luca.leuzzi@cnr.it}

\begin{abstract}

  The critical behavior of the $O(2)$ model on dilute L{\'e}vy graphs
  built on a 2D square lattice is analyzed.  Different qualitative
  cases are probed, varying the exponent $\rho$ governing the
  dependence on the distance of the connectivity probability
  distribution.  The mean-field regime, as well as the long-range and
  short-range non-mean-field regimes are investigated by means of
  high-performance parallel Monte-Carlo numerical simulations running
  on GPU's.  The relationship between the long-range $\rho$ exponent
  and the effective dimension of an equivalent short-range system with
  the same critical behavior is investigated.  Evidence is provided
  for the effective short-range dimension to coincide with the
  spectral dimension of the L{\'e}vy graph for the XY model in the
  mean-field regime.

\end{abstract}

\maketitle

\section{Introduction: interacting models in complex networks}
\label{introduction}
The study of interacting systems defined in complex, non-regular
structures is interesting from at least three points of
view. Statistical mechanical models in graphs are used for the
description of phenomena in different fields, among which one can
cite: 
stock market dynamics,
\cite{PhysRevLett.89.158701,Yang2006377,Yang2008Increasing,Yang2008Critical}
 correlations in
bird flocking, \cite{Bialek2012Statistical} avalanches in brain
activity \cite{DeArcangelis2010Learning} or biological networks.
\cite{DeMartino2012Scalable, DeMartino2012Reaction}
\indent Furthermore, there is a theoretical interest {\it per se} in
the study of criticality in complex networks. The theory of critical
phenomena establishes that the critical properties of systems
interacting in a $d$-dimensional lattice only depend  on the symmetries
of the interaction and on the dimensionality $d$. On the other hand,
when the topology of the interaction is more complicated, e.g.,
translational invariance is lost and symmetries of the lattice are
broken, the dependence of criticality on the structural properties of
the interacting graph is not known in general, although this topic has
been subject of interest since more than two decades and several
results are available for particular models. \cite{Dorogovtsev2008Critical}
\\ \indent A particularly clear case is the spherical model, which has
been proved to be equivalent to the $n\to\infty$ limit of the $O(n)$
model when both models are defined on a lattice.
\cite{Stanley1968Spherical} On general graphs the full equivalence
does not hold anymore, though the critical behavior of the spherical
and $O(n\to\infty)$ still do coincide. \cite{Burioni2000ntoinfty} The
critical properties of the spherical model on a general graph are
exactly known \cite{Cassi1999Spherical} and are such that the
universality class of the transition only depends on a single
quantity: the spectral dimension of the graph, $\bar d$, defined in
terms of the low-frequency spectral density of its adjacency matrix
$\varrho(\omega)\sim\omega^{\bar d/2-1}$. This quantity is also
related to the probability of self-return of a random walker in the
graph, and determines the infrared divergences of a Gaussian field
theory defined on the graph.
\cite{Hattori1987Gaussian,Cassi1999Spherical} Remarkably, the
functional dependence of the spherical model critical quantities on a
graph with spectral dimension $\bar d$ turns out to be the same as the
ones of a short-range model in a hyper-cubic lattice with Euclidean
dimension $\bar d$.  This analogy provides a suggestive, physical
sense to the non-integer dimensions appearing in the context of the
theory of critical phenomena.
\\ \indent The spectral dimension also plays a role in the XY model
criticality, which was proved \cite{Burioni1999Inverse} to exhibit
spontaneous magnetization in the ordered phase in a graph of spectral
dimension $\bar d>2$, and absence \cite{Cassi1992Phase} of spontaneous
magnetization for $\bar d\le 2$. The latter phenomenon being well
known in the two-dimensional (2D) XY model,
\cite{Kosterlitz1973Ordering,Kosterlitz1974Critical,Mermin1966Absence}
which is a particular case of this result.
\\ \indent Further numerical and analytical results for the
criticality of other particular models in graphs are available (for a
review see Ref. \onlinecite{Dorogovtsev2008Critical}). The Ising model
was first studied in Small-World networks,
\cite{Barrat2000Properties,0305-4470-33-47-304,PhysRevE.65.066110,PhysRevE.66.018101,Kwak2007Critical}
in Barabasi-Albert networks,
\cite{Bianconi2002166,Aleksiejuk2002260,Kwak2007Critical} and on
general graphs, \cite{Leone2002Ferromagnetic,Dorogovtsev2002Ising}
where it was found that the universality class depends on the
divergence or finiteness of the second and fourth moments of the
degree distribution. In this way, three different critical regimes may
be discriminated: (i) absence of phase transition, when both second
and fourth moments diverge; (ii) a non-mean-field second order
transition, when the second moment is finite; and (iii) a mean-field
second-order transition, when both moments are finite.
\\ \indent Studies of the Ising model in scale-free networks
\cite{Herrero2004Ising} and in correlated growing-random networks
\cite{Bauer2005Phase,Callaway2001Are} were also performed. In the
latter case a phase transition was found of the Kosterlitz-Thouless
(KT) universality class, different from the mean-field nature of the
transition found in (uncorrelated) scale-free networks. This
difference was argued to have its origin in the sign of the
degree-degree correlations (assortativity-disassortativity) of both
types of
networks. \cite{Bauer2005Phase,Dorogovtsev2003Renormalization} The
Potts model has also been investigated,
\cite{Igloi2002First,Dorogovtsev2004Potts,Khajeh2007BerezinskiiKosterlitzThoulesslike}finding
an infinite-order transition for a divergent second moment of the
degree distribution.
\\ \indent Eventually, also the $O(2)$ XY model has been analyzed. In
the 1D Small-world network, it was argued \cite{Kim2001XY} to exhibit
long-range order for arbitrarily low values of the rewiring
probability (like in the Ising case). For uncorrelated
\cite{Yang2008Critical} and correlated \cite{Kwak2007Critical}
scale-free networks an order-disorder transition is observed for a
sufficiently large value of the degree distribution
exponent. Interestingly, as it happens in the Ising case, in the
correlated scale-free network the transition is non-mean-field, unlike
the uncorrelated case. This difference is again ascribed to the
different nature of the degree-degree correlations in both kinds of
graphs. \cite{Yang2008Critical}
\\ \indent Another piece of the puzzle is provided by the numerical
work carried out by Yang {\em et al.}, \cite{Yang2009Critical} in
which the critical behavior of the XY model is studied in uncorrelated
and correlated {\it random} (rather than scale-free) graphs. In the
first case, i.e., the Erd\"os-R\'enyi graph, the transition is found
to be of the mean-field type, while in a randomly growing network it
is claimed that the occurring transition belongs to the KT
universality class.
 \\ \indent Despite numerous results in this field, a unified picture
 of critical phenomena in graphs is still lacking.  For instance, it
 is not clear under what conditions a relation can be established between criticality
 in graphs with spectral dimension $\bar d$ and short-range models in
 $\bar d$ dimensional lattices, nor what is the
 relation with the conjectured influence of dissasortativity on
 criticality.
 
 \subsection{L\'evy lattice}
 In order to study networks in different universality classes in a
 continuous way, and, thus, deepen the relationship between long-range
 system Euclidean dimension $d$, short-range equivalent Euclidean
 dimension $D$ and spectral dimension $\bar d$ we adopt the so-called
 {\em L\'evy} or long-range {\em dilute} lattice.
 \cite{Leuzzi2008Dilute} It is a graph in which two nodes are
 connected with a probability decaying as a power $\rho$ of their
 distance in a given $d$-dimensional lattice. The total number of
 links in the system is $Nz/2$, $z$ being the average connectivity of
 a node.  While for large enough $\rho$ one recovers the
 $d$-dimensional hyper-cubic lattice, the $\rho=0$ L\'evy graph limit
 corresponds to the random Erd\"os-R\'enyi graph, such that the $zN/2$
 bonds are chosen at random from the set of all $N(N-1)/2$ possible
 bonds.  With respect to the fully connected version of the model, thus,
the study of the model defined on the L\'evy graph allows for
 more efficient computation, since the number of couplings grows only
 linearly with the size $N$.  Varying the power $\rho$, one actually
 acts as if continuously varying the dimension of a $D$-dimensional short-range
 lattice model, equivalent --from the critical behavior point of
 view-- to the long-range model.

The possibility yielded by L{\'e}vy lattices of changing the effective
dimensionality, freely choosing the universality class of the model
without compromising the computational complexity, is useful to
approach different problems: the applicability of the replica symmetry
breaking theory in and out of the spin-glass mean-field
regime,\cite{Leuzzi2008Dilute,Leuzzi2011Bond} the existence of the
Almeida-Thouless critical line above the spin-glass upper critical
dimension in Ising\cite{PhysRevLett.103.267201,Katzgraber2009Study}
and Heisenberg\cite{PhysRevB.84.014428} systems, the criticality of
the 3-spin spin-glass,\cite{Larson2010Numerical} the random field
Ising model transition at zero temperature\cite{Leuzzi2013} and the low
temperature behavior in Heisenberg spin-glasses (including the
spin-chirality decoupling)
\cite{PhysRevLett.105.097206,PhysRevB.83.214405} as well as in
$O(m\to\infty)$ spin-glasses. \cite{Beyer2012Onedimensional}

\subsection{Criticality regimes}

 For fully connected systems with (ordered or disordered) long range
 interactions decaying with the $\rho$-th power of the distance in a
 $d$-dimensional hyper-cubic lattice, three regimes can be identified:
 \begin{itemize}
 \item $d<\rho<\rho_{\rm mf}(d)$, in which the system undergoes a mean-field
   transition;
 \item $\rho_{\rm mf}(d)<\rho<\rho_{\rm sr}(d)$, in which infra-red divergences
   take place, to be dealt with renormalization group;
 \item $\rho>\rho_{\rm sr}(d)$ where the critical behavior is short-range-like.
 \end{itemize}
The value of $\rho_{\rm mf}(d)$ depends on the specific theory and its
symmetries, thus being different in ordered \cite{Fisher1972Critical}
($\rho_{\rm mf}=3d/2$) and disordered \cite{PhysRevB.27.602}
($\rho_{\rm mf}=4d/3$) systems.  The exponent $\rho_{\rm sr}(d)$ is
defined as the value of $\rho$ at which long-range and short-range
two-vertex functions display the same scaling behavior: $\rho_{\rm
  sr}(d)-d=2-\eta_{\rm sr}(d)$, where $\eta_{\rm sr}$ is the anomalous
scaling exponent of the space correlation function in the
$D$-dimensional short-range counterpart.  The above scenario holds on
the L\'evy lattice, as well. Where, besides, the mean-field regime is
found also below $\rho=d$, down to $\rho=0$.
  
  Critical exponents are functions of $\rho$, as
  \begin{equation}
  \eta_\rho=2-\rho+d
\label{eq:eta_rho}
  \end{equation} for any $\rho$ (the $\eta$ long-range exponent
  is not renormalized) and 
  \begin{equation}
  \nu_\rho=(\rho-d)^{-1}
  \end{equation}
  valid only in the mean-field
  regime.  These expressions are formally the same both in
  ordered\cite{Fisher1972Critical} and
  disordered\cite{PhysRevB.27.602,Leuzzi1999Critical} systems, whereas
  different is their dominion in $\rho$ and the renormalized
  expression for $\rho>\rho_{\rm mf}(d)$. The prediction for the
  $\eta_\rho$ exponent has been compared with the outcome of numerical
  simulations in the case of the long-range Ising
  ferromagnet.\cite{Picco2012Critical}

\subsection{Short-range and long-range equivalence conjecture}
Starting from the field-theoretic representation in the free theory
limit, an equivalence
between $\rho$ and $D$ can be conjectured:
\cite{Leuzzi2008Dilute}
\begin{eqnarray}
D=\frac{2d}{\rho-d} & & \rho\in(d:2+d]
\label{eq:Dsr_vs_drho}
\\
D=d \qquad&& \rho\geq 2+d
\nonumber
\end{eqnarray}
   This is exact up to $\rho=\rho_{\rm mf}$ (or down to $D=D_u$)
 but provides a $\rho_{\rm sr}=2+d$, that is wrong.  It can be
 improved as \cite{Larson2010Numerical, Banos2012Correspondence}
 \begin{equation}
 D=\frac{2-\eta_{\rm sr}(D)}{\rho-d}d 
 \label{eq:Dsr_etasr_vs_drho}
 \end{equation}
for which $D=d$ at the right $\rho_{\rm sr}=d+2-\eta_{\rm sr}(d)$. The
above relationships hold in absence of external fields and do not
depend on the specific symmetries of the system, nor on the presence
of any long-range order at all (as in the quenched disordered case).
What changes is the range of values of $\rho$ determining the
universality class to which the model belongs.

 Eq. (\ref{eq:Dsr_etasr_vs_drho}) has been carefully tested in 1D
 L\'evy Ising spin-glasses for $\rho>\rho_{\rm mf}(1)=4/3$, verifying
 that the equivalent short-range critical behaviors are actually
 consistent both for $D=3$ ($\rho=1.792$) and for $D=4$ ($\rho=1.58$),
 but the compatibility is better the higher
 $D$.\cite{Banos2012Correspondence} 
 In 2D (fully connected) ordered
 Ising model at $\rho=1.6546$ and $1.875$ that,  
 according to Eq. (\ref{eq:Dsr_etasr_vs_drho}), should correspond,
 respectively, 2D and 3D \cite{Angelini12} numerical estimates of
 critical exponents are consistent nearer to the mean-field threshold
 (3D) but for $\rho=1.875$ they do not appear compatible anymore with
 the 2D model.  
 These observations hint that
 Eq. (\ref{eq:Dsr_etasr_vs_drho}) is but an approximating
 interpolation beyond mean-field. In the following we will test the conjectured relation 
Eq. (\ref{eq:Dsr_etasr_vs_drho})  on the 2D XY model on L\'evy graph.

To make reading more fluid and avoid notation ambiguities, in
Tab. \ref{dimensions} we summarize various dimensions we refer to
in this work.
\begin{table}[t!]
\begin{center}
\begin{tabular}{|c|c|}
\hline
$d$ & Auxiliary lattice dim. \\
\hline
$D$ & Critically equivalent short-range lattice dim. \\
\hline
$\bar d$ & L\'evy graph spectral dim. \\
\hline
$D_{\rm u}$ & Upper critical dim.\\
\hline
\end{tabular}
\caption{Summary  of the different dimensions considered. }
\label{dimensions}
\end{center}
\end{table}

There are some aspects that should be clarified also in this
context. An elementary question is about the relationship between  the
spectral dimension $\bar d$ of the graph and the short-range
equivalent dimension $D$, cf. Eq. (\ref{eq:Dsr_vs_drho}).  A rigorous
derivation of the critical properties of long-range dilute models as a
function of the power $\rho$ is still lacking, and also an argument
stating under what conditions they are equivalent to the fully
connected case with the same value of the power $\rho$.  In what
follows we will clarify some of the mentioned issues in the $O(2)$
case. Along with theoretical arguments,
we shall present the outcome of numerical simulations run on Graphics
Processing Units (GPUs) with a {\em ad hoc} optimized code, whose
dynamics is based on the Metropolis, Parallel Tempering and
Over-relaxation algorithms, suited to study continuous spin models
interacting in graphs with arbitrary topology (and possibly
randomness).

The paper is organized as follows: in Sec. \ref{considerations} we
provide a theoretical argument to support the existence of three
different
universality classes of the $O(2)$ model defined on a dilute-2D graph
with power $\rho$, relative to three 
intervals of the power $\rho$. We yield numerical evidence in support of the
fact that in the mean-field regime the Euclidean $D(\rho)$ is,
actually, the spectral dimension of the graph with power $\rho$.  In
Sec. \ref{NumericalMethod} numerical methods are exposed.  We present
numerical results in Sec. \ref{NumericalResults} and our conclusions
in Sec.  \ref{Conclusions}. 
  
\section{Criticality of the XY model in the 2D L\'evy graph}
\label{considerations}

We are concerned with the ferromagnetic $O(n)$ model,  defined by the Hamiltonian:

\beq
H=-\sum_{i<j = 1}^N J_{ij}\,\mn S_i \cdot \mn S_j
\label{HXY}
,
\eeq
where $\mn S_i$ denotes the dynamic variable on the $i^{th}$ site of
the graph, an $n$-dimensional vector with unit modulus, and the
product is a $n$-dimensional Euclidean scalar product, the XY model
being the $n=2$ case. The values of the adjacency matrix $J_{ij}$ of
the graph can be either 0 (no connection) or 1. 

We will study the
dilute L\'evy graph, for which two sites $i$ and $j$ are connected
(i.e., the element of the $J_{ij}$ matrix is 1) with a probability
\barray {\cal P}_\rho(J_{ij}) &=&\frac{1}{ Z} |\mn r_i-\mn
r_j|^{-\rho}
\label{Levyi}
\\
\nonumber
Z&=&\sum_{r}r^{-\rho}
\earray
and such that the total number of bonds is independent from $\rho$ and
equal to $2N$ ($z=4$, for periodic boundary conditions). In Eq. (\ref{Levyi}), the vector $\mn r_i$ corresponds
to the position of site $i$ on a square lattice and the probability is
normalized summing over the set of all possible distances between the
sites of the 2D lattice.  Operatively, the set of possible distances
on lattices of linear size $L$ depends on the boundary conditions
chosen for the numerical simulation being periodic (PBC) or free
(FBC). The maximum distance $r_{\rm max}$ will be $[L/2]\sqrt{2}$ for
PBC or $L\sqrt{2}$ for FBC.

\begin{figure}[t!]                        
\hspace*{-.6cm} \includegraphics[width=1.09\columnwidth]{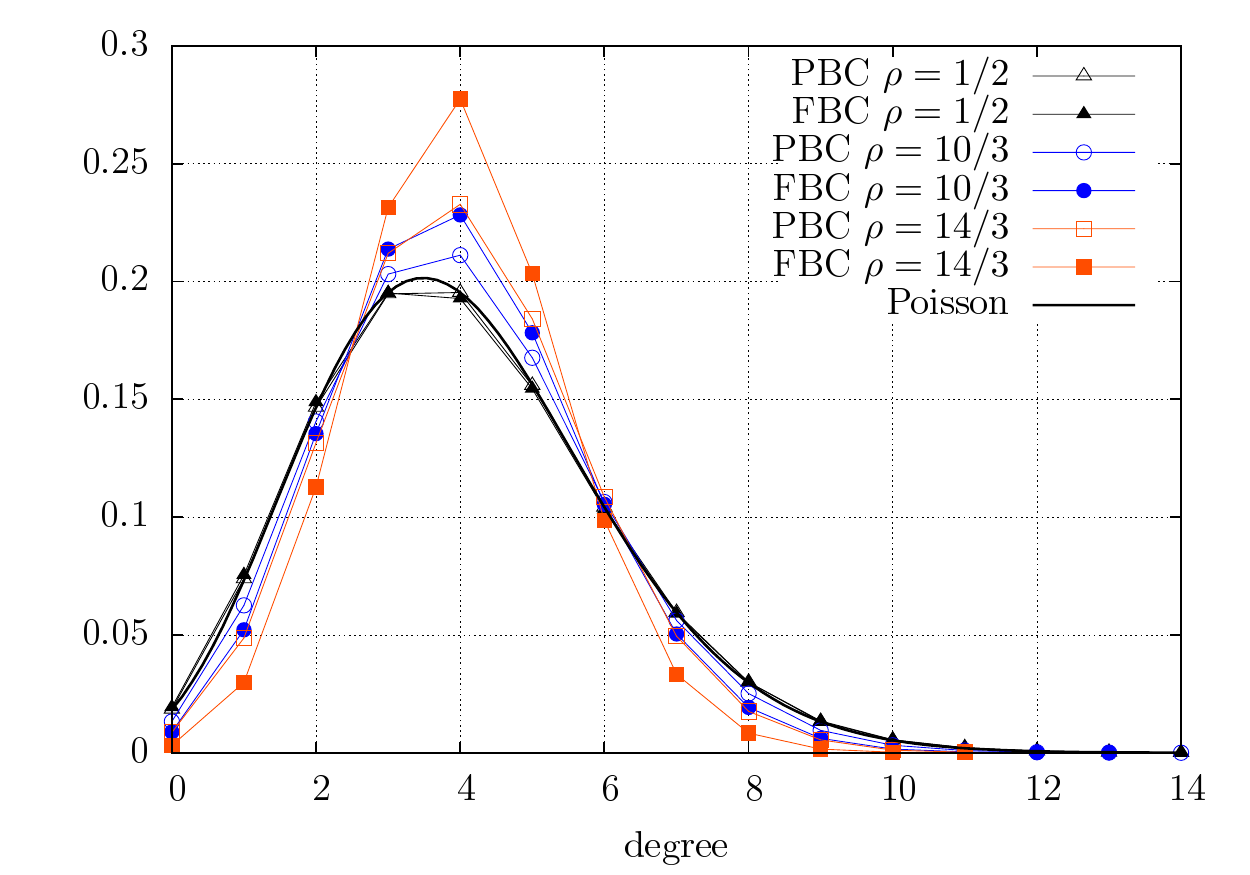} 
\caption{Probability distribution of the degree of connectivity of dilute 2D graphs with $N=256^2$ nodes for three values of the decay exponent $\rho=1/2,10/3$ and $14/3$ of the link probability, one for each  critical regime. Results are reported for both periodic (open symbols) and free (closed symbols) boundary conditions. $\rho=14/3$ is in the short-range regime, $\rho=10/3$ in the non-mean-field regime and $\rho=1/2$ in the mean-field regime. In the latter case the Poisson distribution with average $4$ is displayed for comparison.}
\label{figdegrees}   
\end{figure}

 In Fig. \ref{figdegrees} we show the degree distribution of the
 dilute 2D graph (with free and PBC) for different values of
 $\rho$. While the square lattice limit of the L\'evy graph ($\rho\to\infty$) exhibits a delta function $\delta(z-4)$,
 the $\rho=0$ limit corresponds to the Erd\"os-R\'enyi graph with
 degree distribution given by a Poisson distribution with average
 degree equal to 4. The latter case is independent from the kind of
 boundary conditions.
The differences in the distribution of the number of
  connections per spin in systems  with FBC and with PBC
  in the $\rho>0$ case are finite size effects that are stronger the larger $\rho$.
  Notwithstanding these differences, for what concerns universal quantities (i.e., critical exponents) 
  the outcome of the numerical FSS analysis of the critical behavior
  remains consistent  when FBC are implemented rather than
  PBC, if the simulated sizes are large enough. We will show an instance of this consistency in Sec. \ref{NumericalResults}.
  Unless otherwise stated, however, the results shown in the present paper are obtained from graphs with PBC.

\subsubsection{A dimensional argument}

We now discuss the criticality of the model. Let us first consider the $d$-dimensional fully connected version of our model where each site is connected with any other site and the interaction strength decays with a power-law $J(\mn r)\sim |r|^{-\rho}$ of the distance in a $d$-dimensional lattice. Following Ref. \onlinecite{Fisher1972Critical}, we consider the following effective Ginzburg-Landau Hamiltonian for the long-range model,
a scalar $\phi^4$ Sine-Gordon theory,
\beq
{\cal H}=L^{d} \int \frac{{\rm d}^d{\mn q}}{(2\pi)^d}\,(q^\psi+m^2)\,|\tilde \phi(\mn q)|^2 + \frac{\lambda}{4!}\int {\rm d}^d \mn r\,\phi^4(\mn r)
\label{HGL}
\eeq
where $L$ is the linear size of the system, $\mn q$ is the momentum space index, $m$ is the mass of the theory, $\tilde \phi$ is the Fourier Transform of the scalar field, and $\lambda$ is the coupling strength. The long-range exponent $\psi$ is such that the Fourier Transform of the interaction $J(\mn r)$ goes like

\beq
\tilde J(q) = L^{-d/2}\,\int{\rm d}^d\mn r\,J(\mn r)e^{\imath \mn r \cdot \mn q}\sim q^{-\psi}
\label{def:psi}\eeq
for low $q=|\mn q|$.
In the  long-range fully connected case ($J(\mn r)\sim r^{-\rho}$), it holds $\psi=\rho-d$, and there is a divergence for $\rho=d$, the point at which the number of links, proportional to $\tilde J(0)$, diverges in the thermodynamic limit.

\begin{figure}[t!]           
\begin{center} 
 \includegraphics[width=.99\columnwidth]{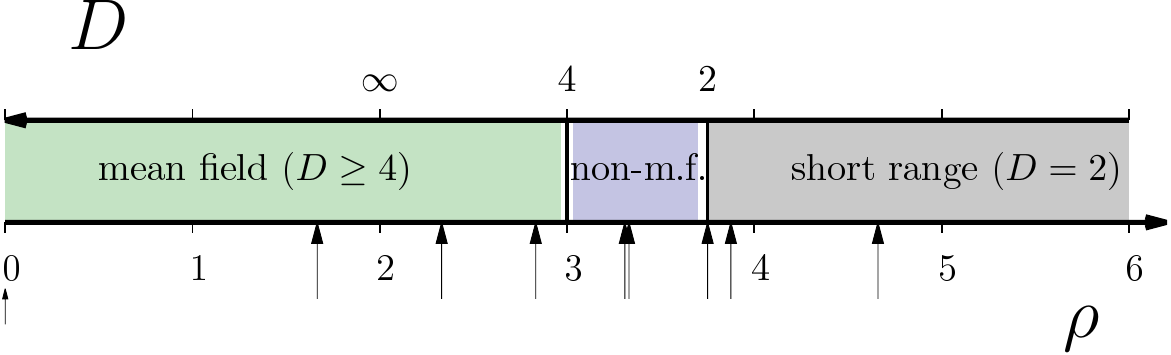} 
\caption{Three domains of $\rho$, relative to the three universality classes of the dilute XY model on the L{\'e}vy graph. The arrows point out at the values of $\rho$ at which simulations have been carried out: $0,1.667, 2.333,2.833, 3.307933,3.333,3.75, 3.875,4.67$.}
\label{figregimes}
\end{center}   
\end{figure}

A dimensional analysis of the Hamiltonian, Eq. (\ref{HGL}), see, e.g., Refs. [\onlinecite{Fisher1972Critical,PhysRevB.27.602,Leuzzi1999Critical,Leuzzi2008Dilute}], shows that the dimension of the coupling constant $\lambda$ is larger than zero whenever $\rho>3d/2$. Below this point, $\lambda$ is an irrelevant variable at criticality:  the system critical behavior is correctly described by a (mean-field) free theory. 
For $\rho>\rho_{\rm rs}(d)$,
the short-range lattice contribution to the propagator $q^2$ takes over the long-range $q^\psi$ contribution
and the Ginzburg-Landau Hamiltonian  corresponds to the one of a $d$-dimensional short-range model. 
Considering the anomalous decay of the correlation function at criticality $\psi$ has to be compared with $2-\eta$ yielding
$\rho_{\rm sr}(d)=2+d-\eta_{\rm sr}(d)$. We will motivate better in the following section such analogy (see also subsection \ref{spectraldimension}).
Eventually, in the fully connected model, an ultra-extensive regime occurs for $\rho<d$, with diverging energy. This is not present in the dilute model, since the number of bonds of the graph is constant. Collecting all the above considerations, we summarize the following dependence of the criticality of the dilute XY model on the exponent $\rho$
(see Fig. \ref{figregimes}):

\begin{enumerate}
\item $\rho \ge \rho_{\rm sr}(d)$: the model should behave similarly to its short-range version in $d$ dimensions, for what concerns criticality, thus belonging to the Kosterlitz-Thouless (KT) universality for $d=2$. This regime will be called the {\it short-range (SR) regime} in the following. 

\item  $\rho \in (\rho_{\rm mf}(d):\rho_{\rm sr}(d))$, with $\rho_{\rm mf}=3d/2$: the system will present a transition different from a KT transition, with exponents different from the mean field-ones, this regime will be denoted as the {\it long-range (LR) non-mean field regime}. 

\item $\rho \le \rho_{\rm mf}(d)$, the system belongs to the mean-field universality class, i.e., its critical properties are the corresponding to a free Gaussian theory in dimension $4$. We will denote this regime as the {\it mean-field (MF) regime}.
\end{enumerate}

\subsubsection{Spectral dimension}
\label{spectraldimension}

Considering Eq. (\ref{eq:Dsr_vs_drho}) one sees
 that the regimes above introduced are in correspondence with the three regimes of an {\it equivalent}  $D$-dimensional lattice model with nearest-neighbor interactions: 
(1)  short-range regime, $D=d$; (2) non-mean-field regime, $D\in (d:D_{\rm u})$; (3) mean-field regime, $D \ge D_{\rm u}$.
 
 Remarkably, this comparison suggests a tight relationship between $D$ and the spectral dimension of the L\'evy graph, $\bar d$. 
 Indeed, in Ref. \onlinecite{Burioni1997Geometrical} it is proved that a fully connected lattice with interaction strength decaying as $r^{-\rho}$ 
 ($r$ being the distance in a $d$-dimensional lattice) has spectral dimension:

\begin{equation}
\bar d= \left\{
\begin{array}{ccl} d \qquad &\textrm{if }& \rho>2+d \\ 
\frac{2d}{\rho-d} \qquad &\textrm{if }& \rho\in(d:2+d]
\end{array}\right.
\label{eqbard}
\end{equation}

It coincides with Eq. (\ref{eq:Dsr_vs_drho}), holding in the mean-field regime $\rho\leq \rho_{\rm mf}(d)=3d/2$.
This implies that the relationship between critical properties of a model on a graph with spectral dimension $\bar d$ and on a lattice of Euclidean dimension $\bar d$,
proved for  spherical and $O(\infty)$ models,\cite{Cassi1999Spherical}
still holds for the $O(2)$ model on L\'evy lattice with $\rho\leq 3d/2$: 
\begin{equation}
D=\bar d
.
\label{SRD_vs_sd}
\end{equation}

\begin{figure}[t!]           
\begin{center} 
 \includegraphics[width=.49\textwidth]{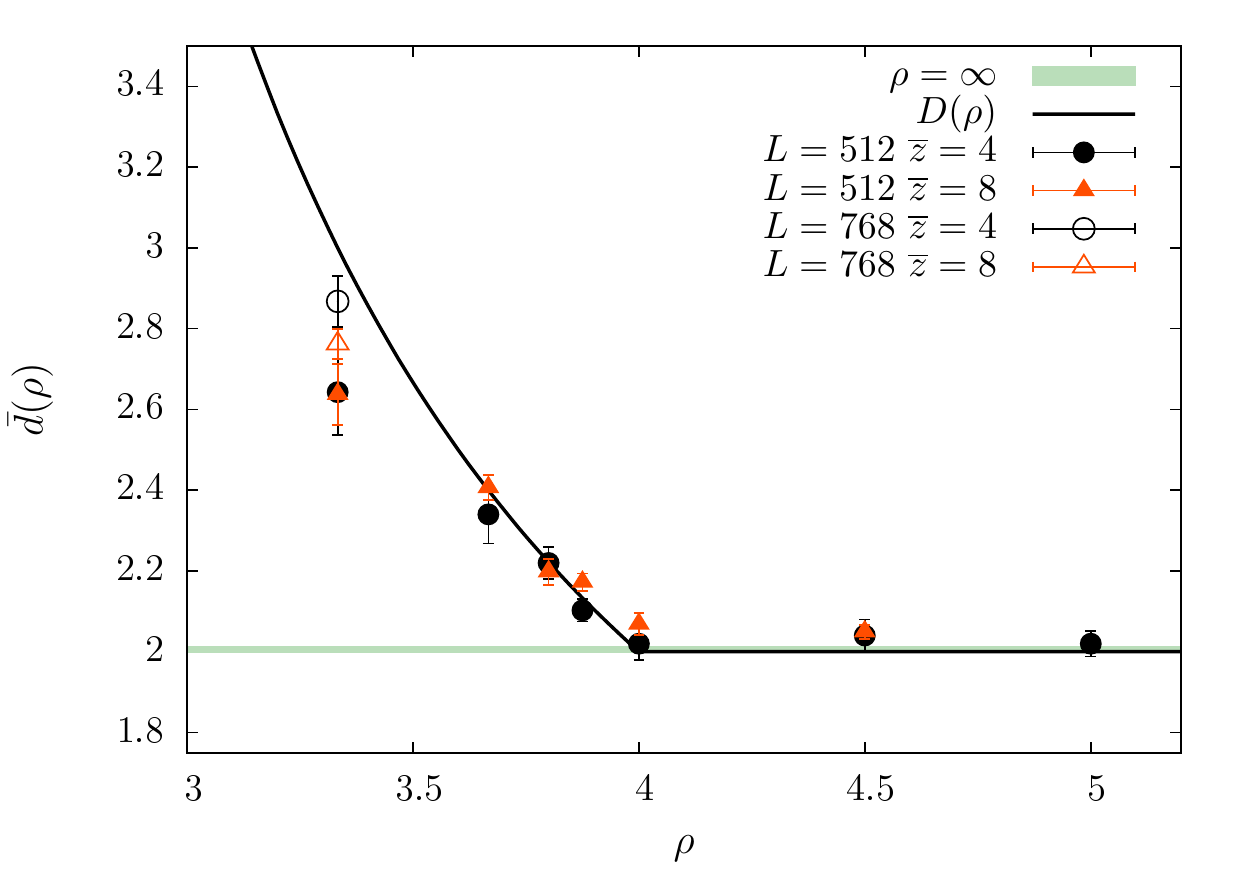} 
\caption{Spectral dimension $\bar d$, cf. Eq. (\ref{eqbard}), or equivalent short-range dimension $D(\rho)$, cf. Eq. (\ref{eq:Dsr_vs_drho}), compared to the numerically estimated
 spectral dimension, versus $\rho$. The light full line is the result for the square lattice case $\rho=\infty$, the black full line is Eq.  (\ref{eqbard}).  
 The numerical estimate has been plotted for two sizes, $L=512$ and $768$ at the smaller $\rho=10/3$ in order to highlight
  relevant finite size effects when long-range connections occur. In all plotted cases, graphs with average connectivity $\overline z=4$ and $8$ have been 
  considered. }
\label{figdrho}
\end{center}   
\end{figure}

In the present section we  provide an analysis of the spectral dimension  directly supporting Eq. (\ref{eqbard}) also in dilute long-range random graphs. We numerically estimated the spectral dimension of 2D dilute L\'evy graphs, with several values of the power $\rho$, through the calculation of the probability of self-return of a random walker in the graph after a time $\tau$, $P(\tau)$, a quantity related\cite{Burioni2005Random} to the spectral dimension via 
\begin{equation}
P(\tau)\sim \tau^{-\bar d/2}
\end{equation}
 for large $\tau$. Our results are summarized in Fig. \ref{figdrho}, in which we compare $\bar d(\rho)$ in Eq.   (\ref{eqbard}) with the estimation of $\bar d$ at the corresponding $\rho$ {\it via} the histogram of the random walker self-return times. As $\rho$ decreases, finite size effects become relevant, e.g., for $\rho=10/3$. Details of the method are reported in App. \ref{App:A}.

Fig. \ref{figdrho} shows that the behavior of the spectral dimension is compatible with Eq. (\ref{eqbard}), even though strong finite size effects take place as $\rho$ decreases towards the mean-field threshold. To get an idea of finite size effects, at $\rho=10/3$ we present results obtained on  graphs  of linear size $L=512$ and $L=796$, with different average coordination number $\overline z$. We observe
 that $\bar d(10/3)$  increases towards the prediction of Eq. (\ref{eqbard}), ${\bar d} = 3$.
This hints that the $\rho$-dependence of the spectral dimension 
of the  L\'evy diluted 2D graph with power $\rho$ coincides with the one of the fully connected version.
Combined with Eq. (\ref{eq:Dsr_vs_drho}), this also  implies the equivalence between the spectral dimension of the graph $\bar d$ and the short-range dimension $D$ of the XY model in 
the mean-field regime, $\rho\leq \rho_{\rm mf}(2)=3$.

\section{Numerical method, algorithm and details of the simulation}
\label{NumericalMethod}
We now expose the numerical method used to analyze the critical properties of the XY model in dilute 2D lattices (\ref{HXY}), via Monte Carlo sampling in finite size realizations of graphs with $N$ vertexes and $2N$ edges.  Given $T$ and $\rho$, we consider both the ensemble average $\<\ldots\>$, at $T$, and the graph (topology) average ${\overline{\phantom {aaa}}}$ at $\rho$:

\beq
\overline{\<O\>}=\frac{1}{{\cal Z}}\sum_{\{J\}}\sum_{\{\mn S\}}O\{\mn S\}\,\exp{\left[-H\{\mn S\}/T\right]}\,{\cal P}_\rho(J )
,
\eeq
where $H$ is the Hamiltonian of the model, Eq. (\ref{HXY}), $O$ is an observable, and ${\cal Z}$ is the partition function. The following quantities are measured: the specific heat

\beq
c = \frac{1}{N}\frac{\partial {\overline{ \<H\>}}}{\partial T} = \frac{1}{NT^2}\left( {\overline{\<H^2\>-\<H\>^2 }} \right)
\label{specificheat}
;
\eeq
the susceptibility 
\begin{equation}
\chi =  N {\overline{\<\mn m^2\>-\< \mn m \>^2}}
\label{chi}
\end{equation}
and the fourth-order Binder cumulant:
\beq
U_4 =  \frac{{\overline{\<\mn m^4\>}}}{{\overline{\<\mn m^2\>^2}}}-1
;
\eeq
where $\mn m$ is the  magnetization: 

\beq
\mn m =  \frac{1}{N}\sum_{j=1}^N\,\mn S_j 
,
\eeq
and where $\{\mn S_j\}_{j=1}^N$ is a given spin configuration. 
Yet another interesting scaling observable is the second moment correlation length $\xi_2$:  \cite{Amit2005Field}
\beq
\xi_2=\frac{1}{2\sin(\mn k_{\rm min}/2)} \left[\frac{\chi(\mn 0)}{\chi(\mn k_{\rm min})}-1\right]^{1/\psi}
\eeq
where $\psi=\rho-d$ in the long-range regime and $\psi=2$ in the short-range regime, cf. Eqs. (\ref{HGL}-\ref{def:psi}),  
$\mn k_{\rm min}=(2\pi/L,0)=(0,2\pi/L)$ is  the smallest momentum in the Fourier space 
 and $\chi(\mn k)$ is the Fourier transform of the  equilibrium  two-point correlation function
\begin{equation}
C(\mn r)= \frac{1}{N}\sum_{\mn i}{\overline{\langle \mn S_{\mn i} \cdot \mn S_{\mn i+\mn r}\rangle}}
\end{equation}
  With these observables we analyze the critical properties of the model around the critical temperature $T_c$ using the scaling relations: 
\barray
U_4(T,N)&=&\tilde U_4(t\,N^{1/\bar\nu}) 
\label{eq:scalingrelations}
\\
c(T,N) &=& \left\{
\begin{array}{ll}
 N^{\alpha/\bar\nu}\, \tilde c(t\,N^{1/\bar\nu}); &
\quad \rho>\rho_{\rm mf}
\vspace*{.3cm}
\\
 \tilde c(t\,N^{1/2}); &
\quad \rho\leq \rho_{\rm mf}
\end{array}
\right.
\label{c_scale}
\\
\chi(T,N) &=& \left\{
\begin{array}{ll}
N^{\gamma/\bar\nu}\, \tilde \chi(t\,N^{1/\bar\nu}) ;
&
\quad  \rho>\rho_{\rm mf}
\vspace*{.3cm}
\\
N^{1/2}\, \tilde \chi(t\,N^{1/2}); &
\quad  \rho\leq \rho_{\rm mf}
\end{array}
\right.
\label{chi_scale}
\\
\xi_2(T,N)&=&\left\{\begin{array}{ll}
L ~\tilde \xi_2(t\,L^{1/\nu_\rho});&
\quad \rho>\rho_{\rm mf}
\vspace*{.3cm}
\\
L^{1/\psi} \tilde \xi_2(t\,N^{1/2});
&
\quad d<\rho\leq\rho_{\rm mf}
\end{array}
\right.
\earray

\noindent where $t=T-T_c$,  $\alpha$ and $\gamma$ are the standard critical exponents and  $\bar \nu$ is the correlation volume exponent. This is
 suited to study scaling relations in graphs and fully connected systems, \cite{Botet1982Size,Kim2001XY} in which the correlation length is no longer well-defined, but for which the correlation volume $V$ diverges at the critical point as $V\sim t^{-\bar \nu}$. The correlation volume exponent is related to the  correlation length exponent of the
 short-range equivalent system by 
\begin{equation}
\bar \nu \equiv \left\{\begin{array}{ll}
D\nu_{\rm sr}(D) &\qquad D<D_u
\\
\vspace*{-.3cm}
\\
D_u\nu_{\rm sr}^{\rm mf}=2&\qquad D\geq D_u
\end{array}\right.
\label{barnu_nusr}
\end{equation}
  According to the conjectured LR-SR equivalence in
 free energy density scaling, \cite{Banos2012Correspondence}  $\nu_{\rm sr}(D) D = \nu_\rho(d) d$, one 
 can hypothesize the following relationship to the LR exponents 
 \begin{equation}
 \bar\nu=\left\{\begin{array}{ll}
 d ~\nu_\rho(d)& \qquad  \rho>\rho_{\rm mf}(d)
 \\
 \vspace*{-.3cm}
 \\
 d~ \nu_{\rho_{\rm mf}}(d)= 2 &\qquad \rho\leq \rho_{\rm mf}(d)=\frac{3}{2}d
 \end{array}\right.
 \label{barnu_nurho}
 \end{equation}
 
 Consequently,  using the Widom scaling relation
\begin{equation}
\frac{\gamma}{\nu_\rho}={2-\eta_\rho} \ , 
\end{equation}
 one has
  \begin{equation}
 \frac{\gamma}{\bar\nu}=\frac{\gamma}{d~ \nu_\rho}=\left\{\begin{array}{ll}
 {\rho}/{d}-1& \qquad  \rho>\rho_{\rm mf}(d)
 \\
 \vspace*{-.3cm}
 \\
{1}/{2} &\qquad \rho\leq \rho_{\rm mf}(d)=\frac{3}{2}d
 \end{array}\right.
 \label{gamma_barnu}
 \end{equation}
 
\noindent that can be easily verified/falsified, thus yielding information about the reliability of the SR-LR equivalence. 

\section{Numerical results}
\label{NumericalResults}

According to the arguments of Sec. \ref{considerations}, and in order to elucidate the nature of the phase transition in each regime, we have run various sets of simulation with the values or $\rho$ reported in the first column of Tab. \ref{criticalexponents}, cf.  Fig. \ref{figregimes},  $\rho=0,1.667$,$ 2.333$,$2.833$, $3.307933$, $3.333$,$3.75$, $3.875$,$4.667$ 
 and on the 2D square lattice, corresponding to $\rho=\infty$.
We have studied finite-size realizations of the system in L\'evy graphs with $N=L^2$, $L=16, 32, 64,128, 256, 384$ nodes.
Each run, for a fixed topology, consists of $2^{21}$ Monte Carlo Steps (MCS). 
We measure observables each 32 MCS. Time averages are performed
on exponentially increasing windows (between $2^k$ and $2^{k+1}$, $k=1,\ldots, 19, 20$).
Topology averages  are performed over a sampling of $N_g$  simulations with 
different realizations of the graph topology, with $N_g$ decreasing for increasing $N$: $N_g=160$ for $L=16, \ldots, 128$, $N_g=6$  for $L=256$, $N_g=4$ for $L=384$. Equilibration checks have been done by comparing time averages
of observables on exponentially increasing  time windows and verifying the consistency of the energy and magnetization histograms and the correlation length values for the last two time windows. As a further equilibration test, we have checked the coincidence of specific heat measurements according to the 
equality in 
Eq.  (\ref{specificheat}).
The algorithm used, a parallel exchange Monte Carlo with Over Relaxation implemented on GPU's, is
described in detail in App. \ref{App:B}.

\begin{figure}[t!]           
\begin{center} 
\includegraphics[width=.99\columnwidth]{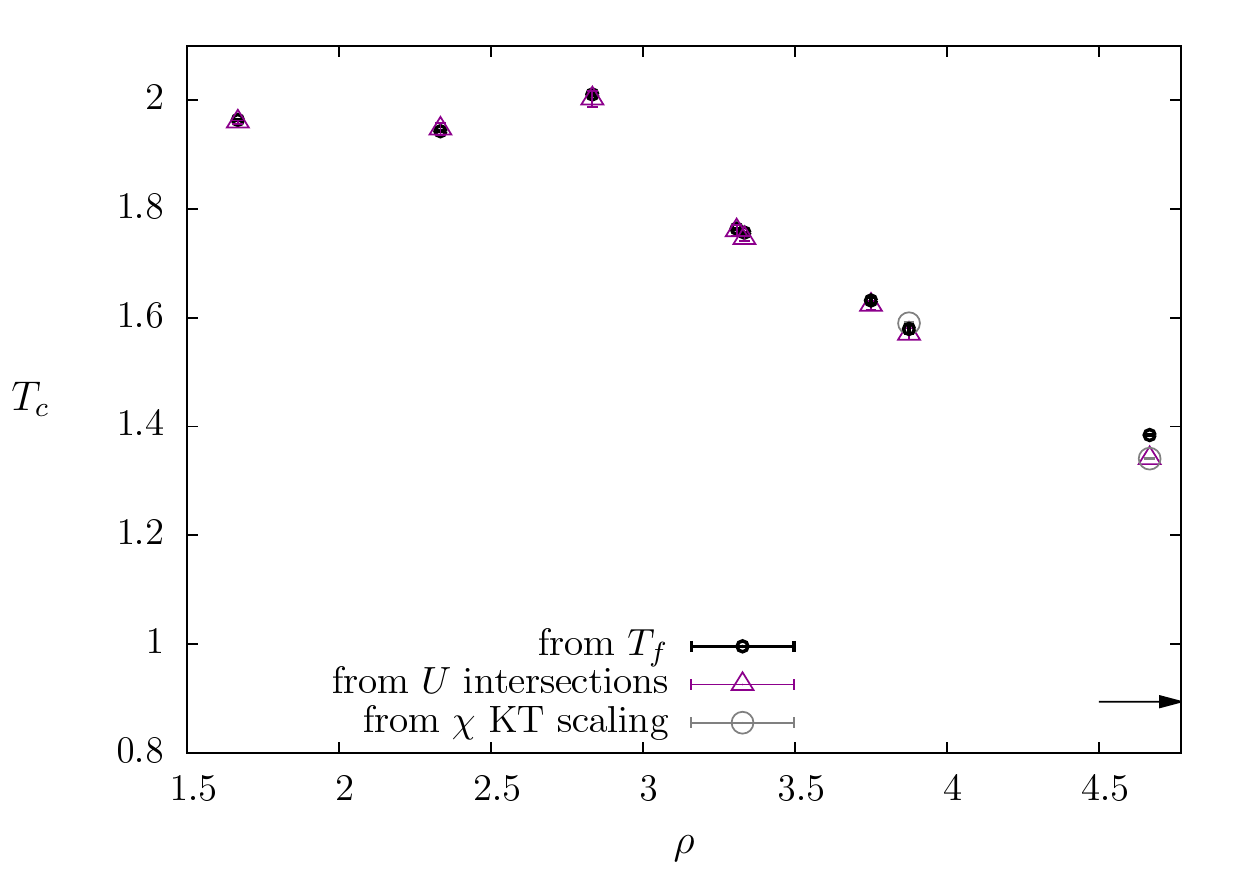}
\caption{Critical temperature in the MF, LR and SR regimes, according to the two different estimation methods described in App. \ref{FSS_multi} and to the $N\to\infty$ extrapolation of the critical temperatures from the KT scaling (\ref{eq:KTscaling}). The horizontal arrow marks the position of the 2D critical temperature.}
\label{fig:Ts}
\end{center}   
\end{figure}

\begin{figure}[t!]           
\begin{center} 
\includegraphics[width=.99\columnwidth]{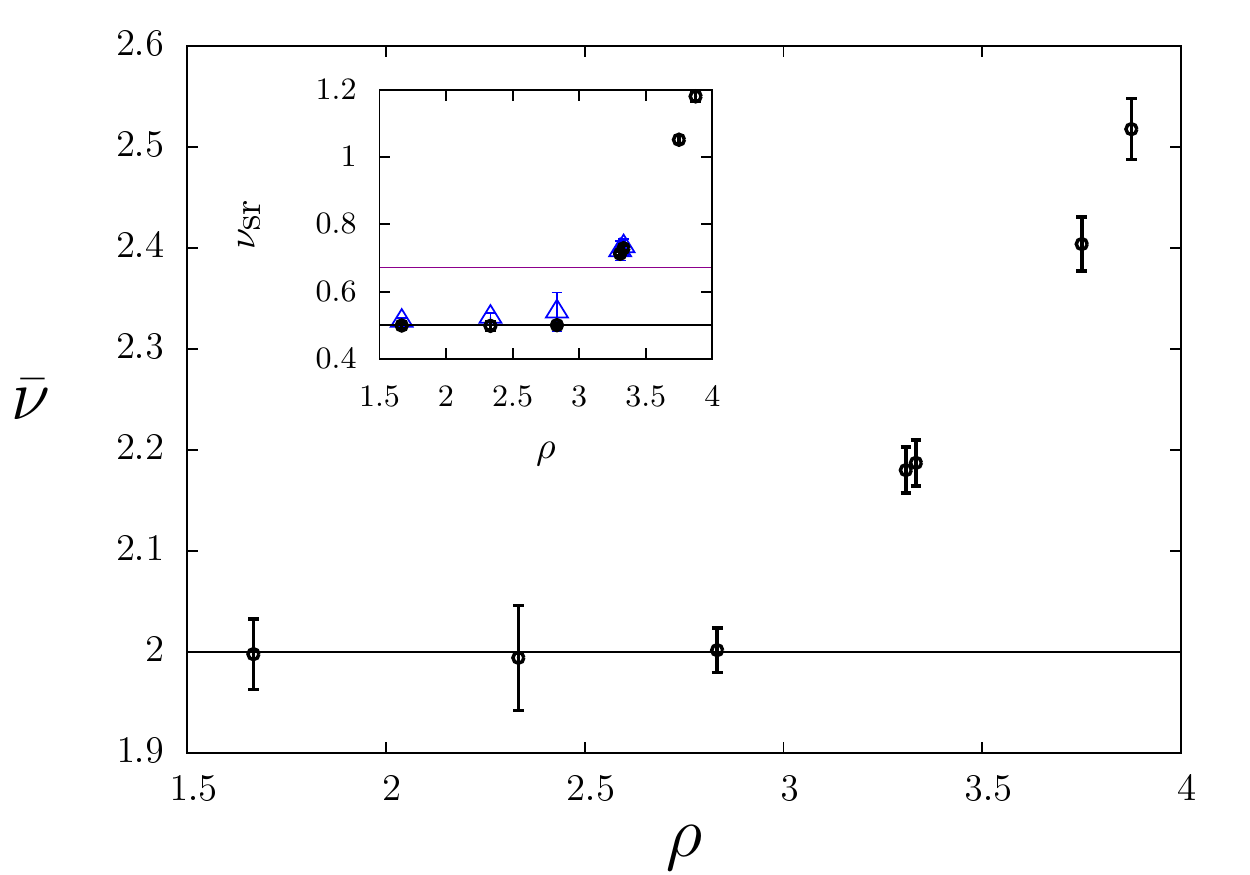}
\caption{Correlation volume exponent $\bar\nu$ vs. $\rho$. Inset: $\nu=\bar\nu/D(\rho)$ with $D(\rho)$ given by (\ref{eq:Dsr_vs_drho}). The horizontal lines are the mean-field value, $1/2$, and the value corresponding to the 3D XY model: \cite{Campostrini2006Critical} $\nu_{\rm 3D}=0.6717(1)$. The blue triangles are the results of an apart analysis ($x^{-1}$ from equation \ref{eq:Tfsscaling}), cf. App. \ref{FSS_multi}.}
\label{fig:nus}
\end{center}   
\end{figure}

\begin{figure}[t!]           
\begin{center} 
\includegraphics[width=.99\columnwidth]{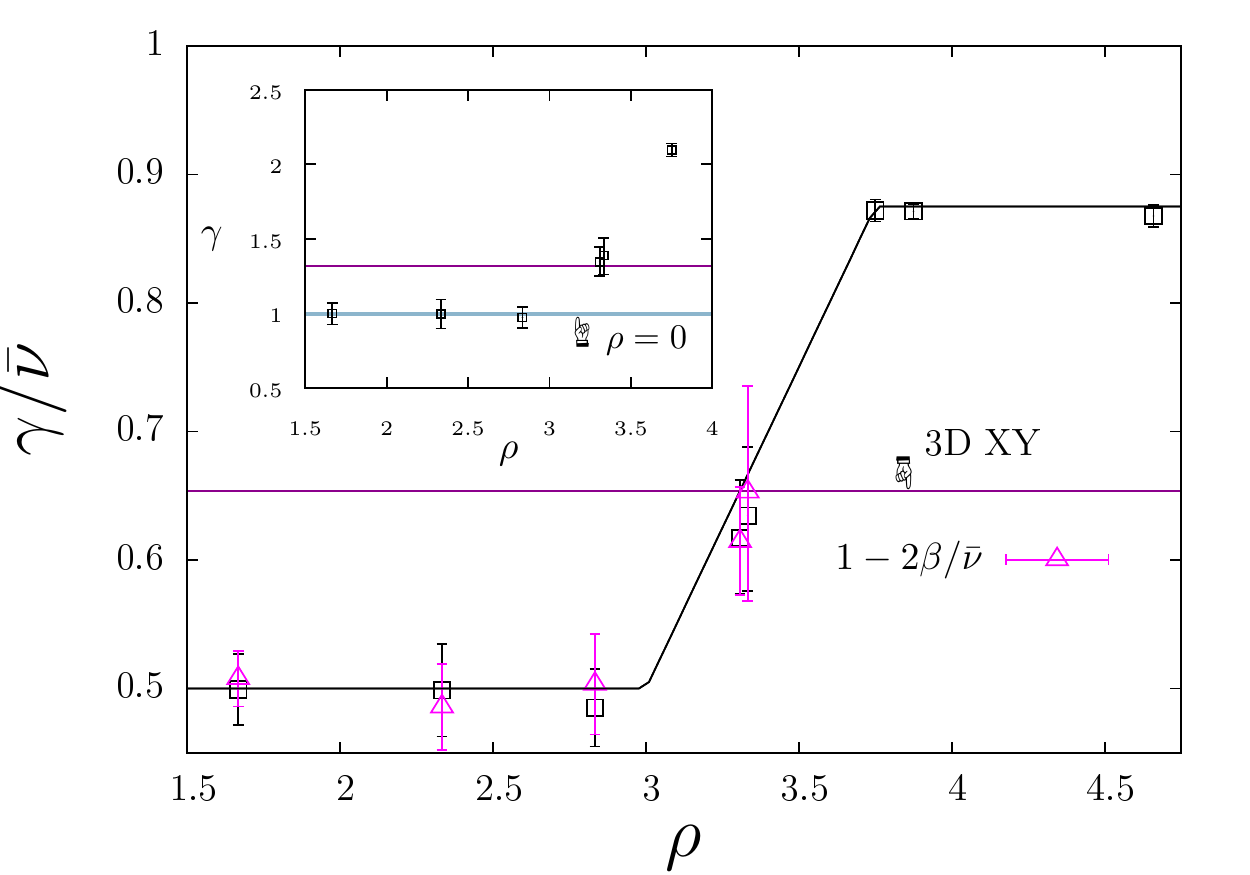}
\caption{Estimations of $\gamma/\bar\nu$ (squares) and of $1-2\beta/\bar\nu$ (triangles). The horizontal line indicates the value of $\gamma_{\rm 3D}/3\nu_{\rm 3D}$ 
\cite{Campostrini2006Critical}, while the black curve is Eq. (\ref{gamma_barnu}). For $3<\rho<4$, the points indicate the average of two values of $\gamma/\bar\nu$ corresponding to two temperatures in the error interval of the estimated critical temperature: $T=1.75$ and $T=1.76$ for $\rho=3.307933$ and $T=1.744$ and $1.75$ for $\rho=10/3$. The inset shows the value of $\gamma$ obtained multiplying the estimation of $\gamma/\bar\nu$ by the estimation of $\bar\nu$ in figure \ref{fig:nus}. The magenta horizontal line indicates $\gamma_{\rm 3D}$, while the green horizontal stripe stands for the estimation of the $\rho=0$ case.}
\label{fig:gammas}
\end{center}   
\end{figure}

In this section we present numerical results supporting the considerations about the XY model criticality as a function of $\rho$ that we presented in section \ref{considerations}. As reference for the rest of the paper, results for the critical behavior are summarized in Figs. \ref{fig:Ts},  \ref{fig:nus}, \ref{fig:gammas} and in Tab. \ref{criticalexponents}. We now proceed to analyze and discuss the critical behavior in the different regimes, starting with the two extreme model limits: the 2D and the Erdos-Renyi XY models.

\begin{widetext}

\begin{table}[t!]
\begin{center}
{\footnotesize{
\begin{tabular}{|c||c|c|c|c||c|c||c|c|c|c|}
\hline
$\rho$ 		& $0$  		& $5/3$       & $7/3$    &$17/6$  &  $3.307933$ & $10/3$ & $3.75$ & $3.875$ &  $14/3$  & $\infty$ \\
\hline
$T_c$ from $T_f$ 	& 1.93(1)  	& $1.96(1)$ & $1.94(1)$ & $2.01(1)$  & $1.76(1)$ & $1.75(2)$  & $1.63(1)$ & 1.58(1) & $1.36(1)$  & 0.89(1) \\
\hline
$T_c$ from $U$ crosses 	& 1.93(1)     	& $1.96(1)$ & $1.94(1)$ & $2.00(1)$  & $1.76(1)$ & $1.75(2)$  & $1.62(2)$ & 1.57(1) & $1.38(2)$  & - \\
\hline
$T_c$ from KT  law  	& -  	& - & - & -  & - & -  & - & 1.59(1) & $1.34(2)$  & $0.893(4)$ \\
\hline
$T_c$ from $\eta$ FSS   	& -  	& - & - & -  & - & -  & 1.60(2) & 1.56(1) & $1.37(1)$  & $0.894(5)$ \\
\hline
$\bar \nu$ 	& $2.00(2)$ & 2.00(3) & 2.00(3) & 2.00(2) & 2.18(2) & 2.19(2) & 2.40(3)  & - & - & - \\
\hline
$\gamma$ &$1.00(1)$& 1.00(4) & 0.99(6) & 0.97(4) & 1.42(7) & 1.45(5) & 2.10(4) & - &  - & -\\ 
\hline
$\beta/\bar\nu$ & 0.25(1) & 0.25(1) & 0.26(2) & 0.25(2) &   0.178(6) & 0.15(2) &- & - & - & - \\
\hline
$\eta$ & - & -  & - & - & -  & - &  0.25(2) & 0.25(1) & 0.26(2) & 0.250(1)  \\

\hline
\end{tabular}
}}
 \caption{Estimations of the critical temperature and of the $\bar\nu$, $\gamma$, $\beta$ and $\eta$ exponents. 
 } 
\label{criticalexponents}
\end{center}
\end{table}

\end{widetext}

\subsection{Kosterlitz-Thouless transition: 2D square lattice limit}
\label{benchmark}

As a benchmark test we have analyzed the outcome of our algorithm in the square lattice (the $\rho=\infty$ limit), where the XY model is known to undergo an infinitely high order  transition, the KT transition. \cite{Kosterlitz1973Ordering,Kosterlitz1974Critical}
The paramagnetic high temperature phase, in which vortices {\it unbound}, displays  exponentially decaying  spatial correlations. The low temperature spin-wave phase is made of coupled pairs of vortices of opposite chirality. 
It is characterized by the absence of spontaneous magnetization and a power-law asymptotic decay of the spatial correlation function. 
              
We can compare our results with the analysis of Ref. [\onlinecite{Gupta1988Phase}], finding excellent agreement for  all the analyzed quantities  ($\chi$, $\<H\>$, $\xi_2$, $c$). Looking at the FSS of the temperatures relative to fixed values of the Binder cumulant, we find  $T_c=0.89(1)$. We also find $T_c=0.894(5)$ with an independent estimate (see below). As a further test, we have looked at several properties of the KT transition, that we will also consider as fingerprints for the classification  that we will do in the following, cf. Sec. \ref{SRRegime}.  We  enumerate them for clarity:

{\it (1)}.  {\em Scale invariance in the whole range $T\le T_c$}. $\qquad$ The scaling functions $\xi_2/L$ and $U_4$ are scale invariant for all  $T<T_c$ in the large-$N$ limit. \cite{Ballesteros2000Critical} Differently from what happens, for example, in standard second-order phase transitions, one observes a superposition of the  finite-size $\xi_2/L$ curves corresponding to different, sufficiently large, values of $L$ (see Fig. \ref{figchi2D}, right panel). In particular, this does not allow to infer the critical temperature from the FSS of crossing points of $U_4$ or $\xi_2/L$. We can, however, estimate $T_c$ by means of the fixed $U_4$ FSS method described in App. \ref{FSS_multi}, yielding $T_c=0.89(1)$. 

\begin{figure}[t!]                        
\begin{center} 
 \includegraphics[width=.49\textwidth]{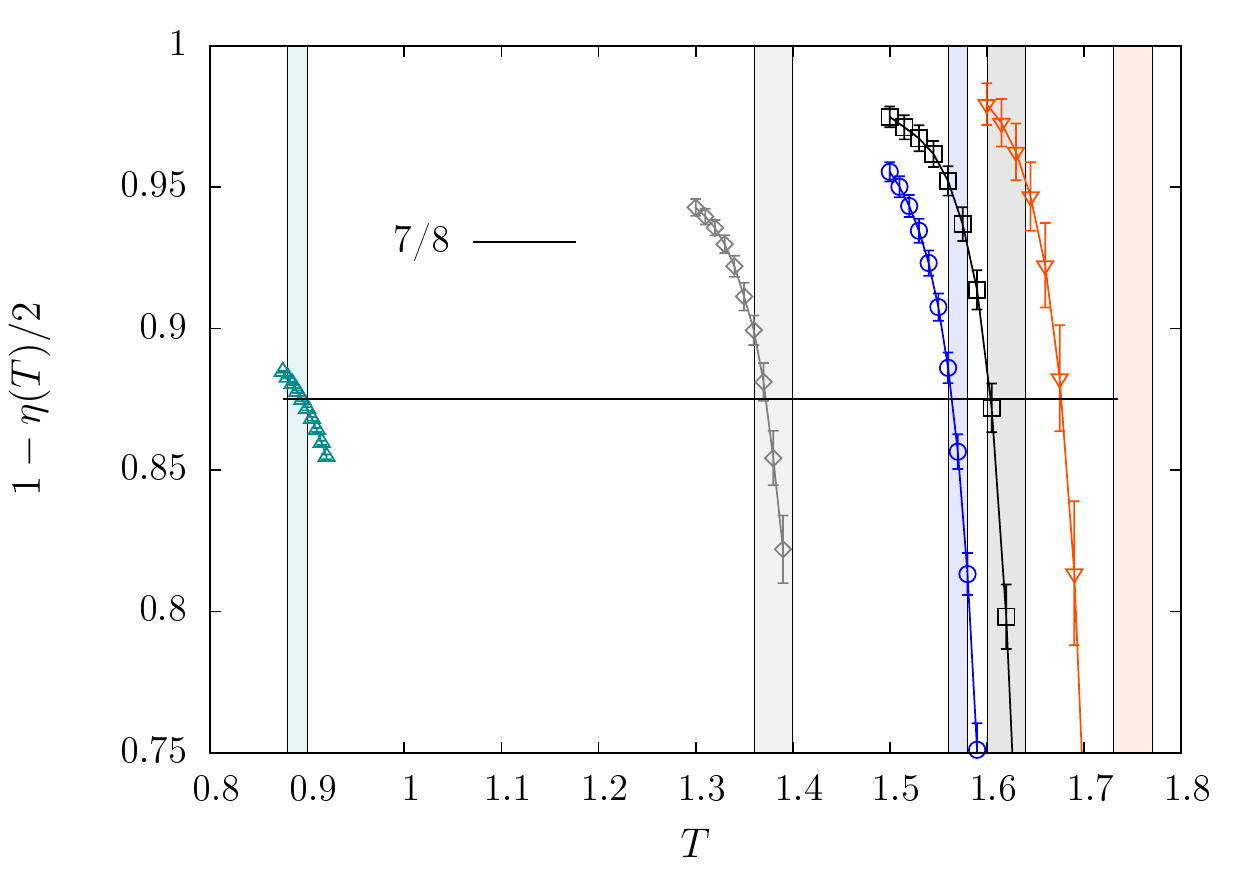} 
\caption{The critical exponent $1-\eta/2$ versus temperature for ---from left to right--- the 2D model, $\rho=4.667$, $3.875$, $\rho_{\rm sr}=3.75$
 and $\rho=3.333$, the latter being in the LR regime. The horizontal line is the value in the SR universality class. The vertical bars are the estimates
 of the critical temperature for the different cases with their error bars. In all  the cases with $\rho\geq 3.75$ the SR critical value of the eta exponent is compatible with our independent estimates of the critical point. In the latter case, not belonging to the SR universality class, this is clearly not occurring.}
\label{eta_T_SR}
\end{center}   
\end{figure}

 {\it (2)}. {\em Susceptibility scaling at criticality.} The susceptibility Eq. (\ref{chi}) at $T=T_c$
is predicted to behave like 
\begin{equation}
\chi \propto L^{2-\eta}\left(\ln L\right)^{-2r} 
\label{chi_SR}
\end{equation}
with $\eta=1/4$ and $r=1/16$.\cite{Amit1999Renormalisation}
Numerically interpolating Eq. (\ref{chi_SR}) at different temperatures we found that the temperature at which $\eta=1/4$
 is compatible, in the statistical error, with our estimate for $T_c$.
 In Fig. \ref{eta_T_SR}) (left most curve) we plot $1-\eta(T)/2$ as estimated in this way. It can be graphically seen that the KT value at criticality is, indeed, reached in the right $T_c$ interval. 
 We can, vice-versa, interpolate a value of $T_c$ as the temperature at which
$\eta(T_c)=1/4$, yielding $T_c=0.894(5)$ with our data, in agreement with recent very precise simulations. \cite{Hasenbusch2005Twodimensional}

{\it (3)}. {\em Absence of magnetization in the low $T$ phase}.  $\qquad$  We find  numerical evidence of vanishing magnetization in the cold phase by looking at the FSS of $\langle \mn m^2\rangle$, or, more precisely,
at $\chi/N=\langle \mn m^2\rangle-\langle \mn m\rangle^2$. The latter term $\langle \mn m\rangle^2$ is numerically not strictly zero but turns out to be always much smaller than  $\langle \mn m^2\rangle$ below the critical point and tends to decrease for the largest sizes.
The FSS behavior of $\chi$ and $\langle \mn m^2\rangle $ below criticality, thus turns out to be practically  identical at the leading term: 
    \begin{equation}
    \sqrt{\langle \mn m^2\rangle} \simeq \sqrt{\chi/N} = {\rm const }\, N^{-\eta/4}
    \label{eq:m2_FSS_2D}
    \end{equation}
    We verified the goodness of this interpolation tending to zero for $N\to \infty$, compatibly with the observation of zero magnetization \cite{Hasenbusch2005Twodimensional}  characteristic of the KT transition. We notice that the  dependence in $N$ is  very slow 
 because $\eta/4=1/16$ in the short-range universality class.

\begin{figure}[b!]                        
\begin{center} 
 \includegraphics[width=.99\columnwidth]{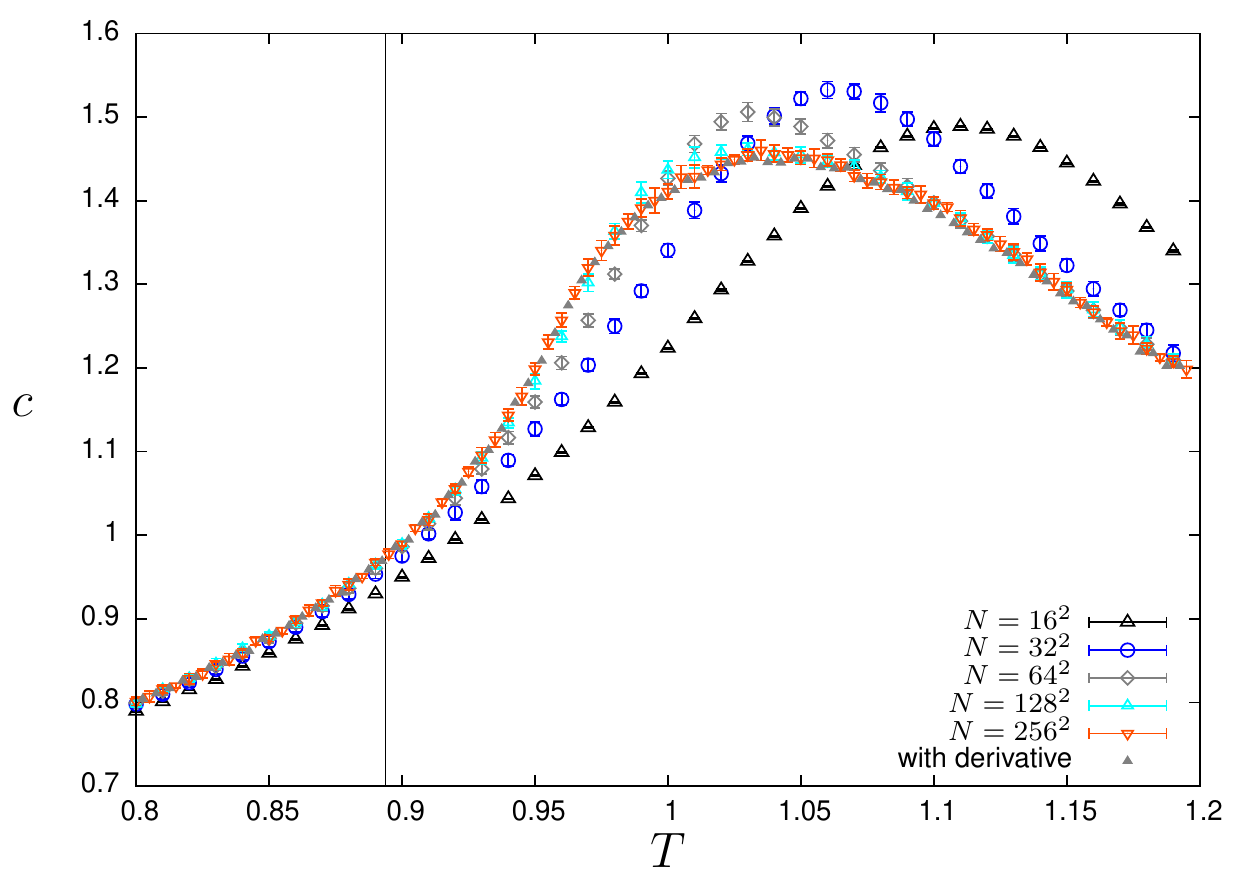} 
\caption{Specific heat versus temperature for the 2D system in lattices of size $N=2^{2n}$, $n=4,\ldots,8$. The vertical line indicates the estimated value of the critical temperature.}
\label{fig2DcV}
\end{center}   
\end{figure}

\begin{widetext}

\begin{figure}[t!]                        
\begin{center} 
 \includegraphics[width=.99\textwidth]{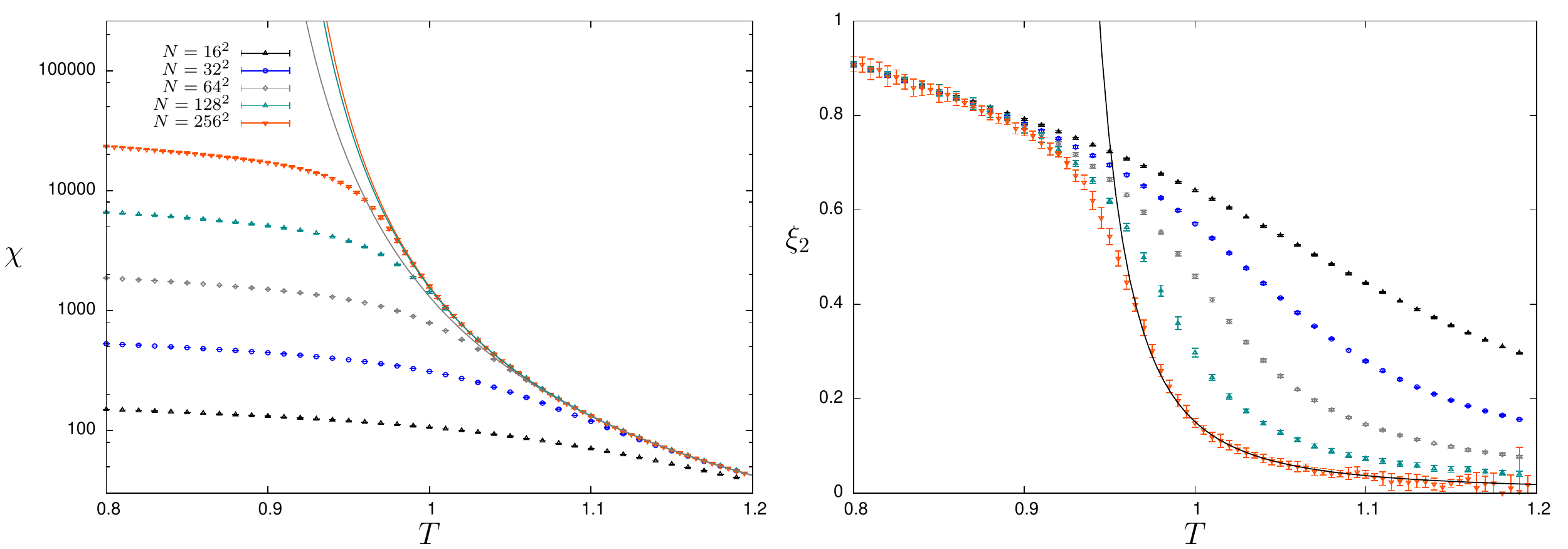} 
\caption{Magnetic susceptibility (left) and correlation length (right) of the 2D model vs. $T$. As $N$ increases, $\chi(T)$ becomes more and more similar to the functional law Eq. (\ref{eq:KTscaling}). In the low-temperature phase the $\xi_2$ curves collapse onto each other.
}
\label{figchi2D}
\end{center}   
\end{figure}  

\end{widetext}

 {\it (4)}.  {\em Continuous specific heat  at the transition}, as can be seen in figure \ref{fig2DcV}.

{\it (5)}. {\em Exponential divergence: Kosterlitz-Thouless law}. $\qquad$
 Susceptibility and correlation length behave at criticality according to the law \cite{Kosterlitz1974Critical}
 \begin{equation}
 X = X_0 \exp \left\{\frac{b_X}{\sqrt{T-T_X}}\right\}
 \label{eq:KTscaling}
 \end{equation}
  with  $X= \xi_2, \chi$. Fig. \ref{figchi2D}, illustrates  FS behavior of $\chi $ and $\xi$.  As the size increases the behavior tends to   Eq. (\ref{eq:KTscaling}) for a larger interval of  temperature. 
 Interpolating $\chi(T,N)$

\newpage

\noindent  with Eq. (\ref{eq:KTscaling}) at fixed $N$ with data in the high $T$ phase we obtain 
 different size dependent curves with parameters $\chi_0(N)$, $b_\chi(N)$ and $T_\chi(N)$. 
 A further estimate of the critical temperature may then be obtained by extrapolating $T_c=T_\chi(\infty)$ 
 with the law $T_\chi(N)=T_\chi(\infty)+{\rm const}\,N^{-3/2}$. With such a method we obtain
  the critical temperature $T_c=0.893(4)$, compatible with the other estimates.

\subsection{Erd\"os-R\'enyi limit}

As a further check we have  studied the $\rho=0$ limit,  corresponding to a random Erd\"os-R\'enyi graph with a Poisson distribution of the degree of connectivity and average degree equal to $4$. We report 
  numerical evidence for a second-order  mean-field  phase transition. Our results are compatible with the theoretical values of the critical exponents $\bar \nu=2$, $\gamma=1$ and $\alpha=0$, as argued in section \ref{considerations}, and are in agreement with the numerical estimates of Refs. 
[\onlinecite{Kim2001XY,Yang2009Critical}], where the mean-field transition on Erd\"os-R\'enyi graphs of average degree $3$ and $8$, respectively, have been analyzed. 

As in the previous subsection, we now present the salient analysis for the Erd\"os-R\'enyi case. The critical temperature is 
estimated from the FSS of the value $T(N)$ at which the cumulant $U_4(T,N)$ intersects with $U_4(T,N/4)$.
Assuming the FSS $T(N)=C N^{-1/\bar \nu}+T_{c}$, we obtain $T_c=1.93(1)$, in agreement with the analytic value $T_c=1.9361$.\cite{MoronePrivate} 
This is also the point at which the specific heat curves for different sizes cross each other, according to the scaling 
law Eq. (\ref{c_scale}). 

The exponent $\bar \nu$ may be estimated by interpolating the relation
\begin{equation}
\frac{\partial U_4(T,N)}{\partial T}\biggr|_{U_4={\rm const}} \sim N^{1/\bar \nu}
 \label{U4dot}
 \end{equation}
 where the derivative of $U_4$ with respect to the temperature is evaluated at fixed values of $U_4$ in the scaling critical 
 region, yielding
   $\bar \nu=2.00(2)$, in agreement with the mean-field value $\bar\nu=2$. 
 Further numerical estimates for the mean-field  critical exponents may be found in the rescaling of the functions
 $\chi$ and $U_4$ as shown for qualitatively similar mean-field cases (see later).
We remark that, although the values of the critical exponents are the mean field ones, the value of the critical temperature is not a universal quantity and does not coincide with the Gaussian mean-field value\cite{Hattori1987Gaussian}  $T=2$. 

\begin{figure}[b!]                        
\begin{center} 
 \includegraphics[width=.49\textwidth]{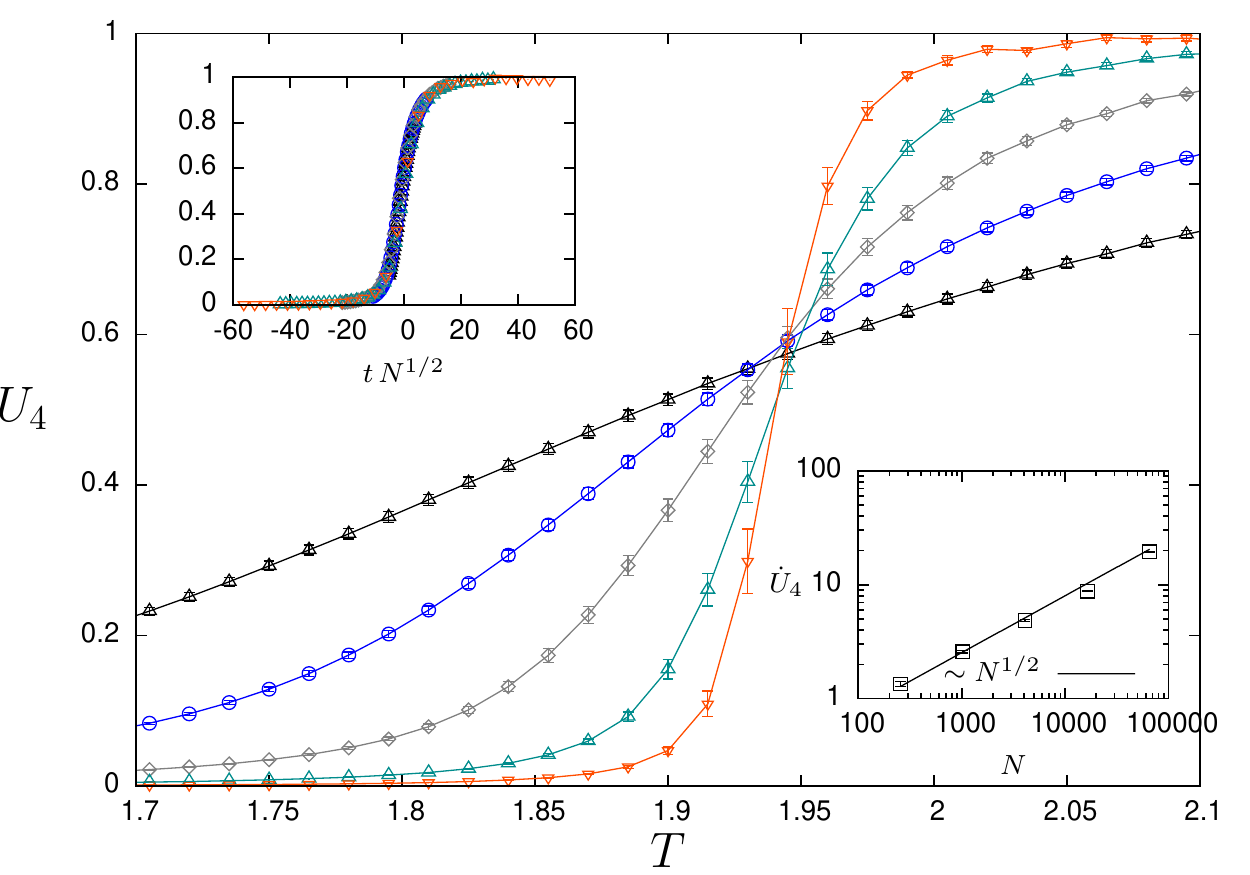} 
\caption{Binder cumulant versus temperature for $\rho=2.333$ in the mean-field regime. The upper inset shows the scaling (\ref{eq:scalingrelations}) with $\bar\nu=2.00(3)$ and the lower inset is the calculation of $\bar\nu$ with the relation (\ref{U4dot}).}
\label{figrhomf1U4}
\end{center}   
\end{figure}

We also checked the Rushbrooke scaling relation $2\beta+\gamma=2-\alpha$ by observing the right scaling of 
the

\begin{widetext}

\begin{figure}[t!]                        
\begin{center} 
\vskip 3.8 mm
 \includegraphics[width=.99\textwidth]{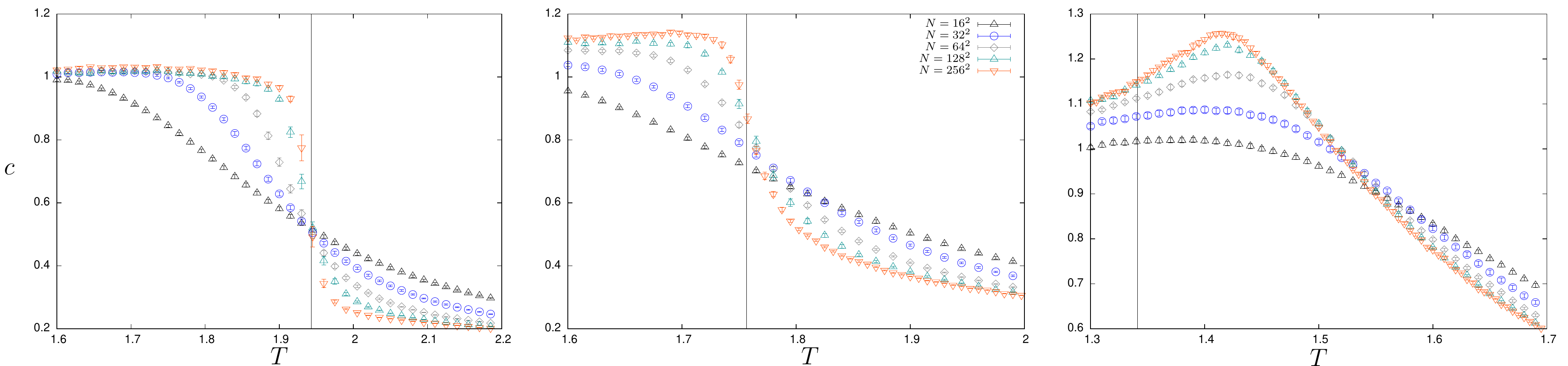} 
\caption{Specific heat versus temperature for different L\'evy lattices.
Left: dilute model with $\rho=7/3$ in the mean-field regime. Center: dilute model with $\rho=10/3$ in the non-mean-field long-range regime. Right: dilute model with $\rho=14/3$ in the short-range regime. In the last case the large $N$ limit of $c$ is regular and continuous at any $T$.}
\label{figcV}
\end{center}   
\end{figure}

\begin{figure}[t!]                        
\begin{center} 
\vskip .4 cm
 \includegraphics[width=.99\textwidth]{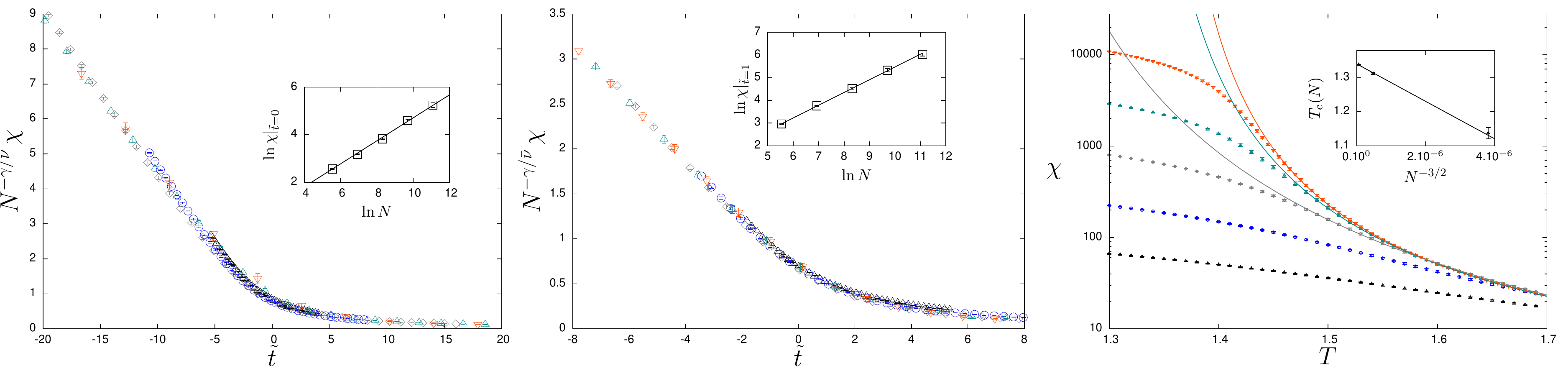} 
\caption{Magnetic susceptibility versus $T$ for different L\'evy graphs Left: Susceptibility scaling, according to Eq. (\ref{eq:scalingrelations}), in the mean-field regime, $\rho=7/3$, $T_c=1.94(1)$, $\bar\nu=2.00(3)$. Inset: finite size behavior of $\ln \chi$ at fixed $\tilde t$, yielding $\gamma=0.99(6)$.
Center: Susceptibility scaling for $\rho=10/3$, in the non-mean-field long-range regime; $T_c=1.75(2)$,  $\bar \nu=2.19(2)$.  Inset: FSS  of $\ln \chi$ at fixed $\tilde t$, yielding  $\gamma=1.45(5)$.
Right: Susceptibility vs. temperature for the L\'evy short-range case $\rho=14/3$. 
As the size increases, $\chi(T)$ becomes more and more similar to the functional law Eq. (\ref{eq:KTscaling}), with which we fitted the data of the systems of size $N=64^2,128^2$ and $256^2$  in a temperature interval beginning at $T=1.48$, 1.47, 1.45 respectively. The $N$-dependence of the so obtained $T_c$ is shown in the inset, together with the fit which extrapolates to $N=\infty$.
}
\label{figchi}
\end{center}   
\end{figure}  
              
\end{widetext}
\newpage

\noindent magnetization with the mean-field exponent $\beta=1/2$.
Indeed, an important  feature of this mean-field transition is that the low-temperature phase presents a finite magnetization, 
and this is confirmed by  FSS analysis of ${\overline{\<\mn m^2\>}}$ in the low $T$ phase.

\subsection{Long-range mean-field regime}
\label{meanfieldregime}

We repeated the same analysis for $\rho=5/3$, $7/3$ and $17/6$, in the mean-field regime.
The first value is nearer to the limit of convergence ($\rho=d=2$) of the fully connected model, where the largest differences with the dilute model could possibly  arise.
The last value of $\rho$ is very near the mean-field threshold  $\rho_{\rm mf}=3d/2=3$.
 Through  FSS of the Binder cumulant we estimate  $T_c=1.96(1),1.94(1)$ and 2.01(1), respectively. In Fig. \ref{figrhomf1U4}
we show the Binder cumulant and its scaling for $\rho=7/3$. 
The derivative of $U_4$ with respect to $T$  allows to estimate
 $\bar \nu=2.00(3),2.00(3)$ and $2.00(2)$, respectively. These are all compatible with mean-field theory, cf. Eqs. (\ref{barnu_nusr})-(\ref{barnu_nurho}).

From the data reported in Fig. \ref{figcV}, left panel, one observes a scaling of the type $c(T,N)=\tilde  c(t\,N^{1/2})$, suggesting that $\alpha=0$. We checked the $\gamma=\bar \nu/2$ mean-field scaling relation for the exponents by fitting the function
 $\ln \chi(T,N)=\gamma/\bar\nu \ln N + \ln \tilde \chi(\tilde t)$ as a function of $\ln N$ for fixed values of $\tilde t$ in 
 the scaling regime obtaining $\gamma=1.00(4),0.99(6)$ and 0.97(4), respectively. The rescaled $\chi$ curve
 for $\rho=7/3$ is plotted in  Fig. \ref{figchi}, left panel.

\begin{figure}[t!]                        
\begin{center} 
 \includegraphics[width=.49\textwidth]{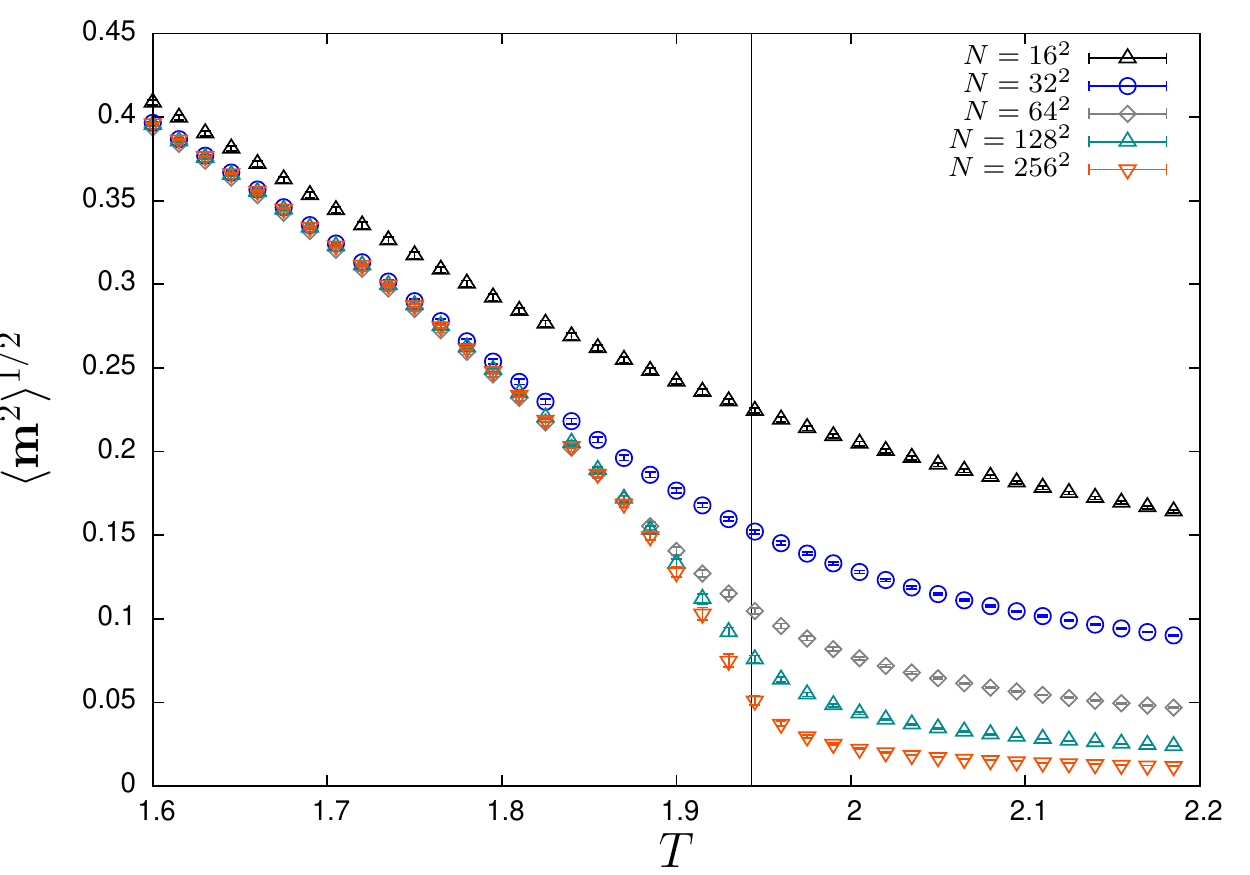} 
\caption{Square root of the average squared magnetization versus temperature for $\rho=2.333$, in the mean-field  regime. As $N$ increases, the low temperature phase exhibits finite spontaneous magnetization.}
\label{figrhomf1absmag}
\end{center}   
\end{figure}

Finally, one finds  that there is spontaneous magnetization in the low-temperature phase. 
The size dependence 
of the square root of ${\overline {\< \mn m^2\>}}$ is very poor and practically no finite size scaling is observed at the largest simulated sizes. 
In Fig. \ref{figrhomf1absmag} the plot for $\rho=7/3$ is shown (left panel).

These numerical results strongly hint that the system belongs to the mean-field universality class in the  range $\rho\in [0:3]$. In the whole range the critical exponents are well defined
and estimated.

\subsection{Long-range non-mean-field regime}

\begin{figure}[t!]                        
\begin{center} 
 \includegraphics[width=.49\textwidth]{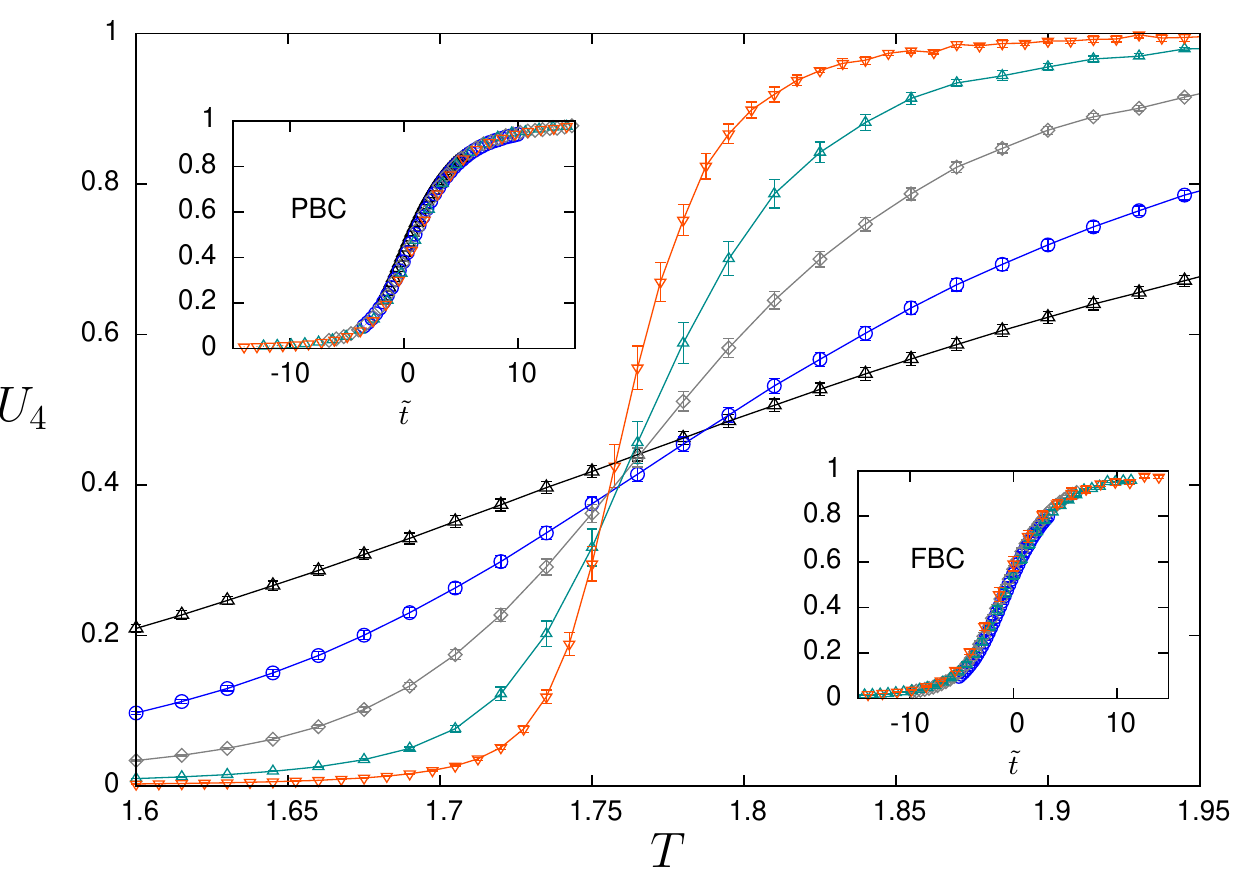} 
 \caption{Binder cumulant versus temperature for $\rho=10/3$ in the non-mean-field regime. The upper inset shows the scaling as in figure \ref{figrhomf1U4} with $\bar\nu=2.19(2)$, while the lower inset shows the scaling int the $\rho=10/3$ system with FBC and with $\bar\nu=2.24(4)$.}
\label{figrhonmfU4}
\end{center}   
\end{figure}

An analogous investigation leads to a different behavior in the non-mean-field regime, for $3<\rho<3.75$. We simulated systems at $\rho=3.3079$ and $3.3333$. 
The critical temperature estimates obtained from the $U_4$ crossing points (cf.,  Fig. \ref{figrhonmfU4} for $\rho=3.3333$) are, respectively,
   $T_c=1.76(1)$ and $1.75(2)$. The FSS analysis of Eq. (\ref{U4dot}) reveals a correlation volume exponent larger that the mean-field value $2$ and increasing with $\rho$:   $\bar \nu=2.18(2)$ and $2.19(2)$.
   
\begin{figure}[t!]                        
\begin{center} 
 \includegraphics[width=.49\textwidth]{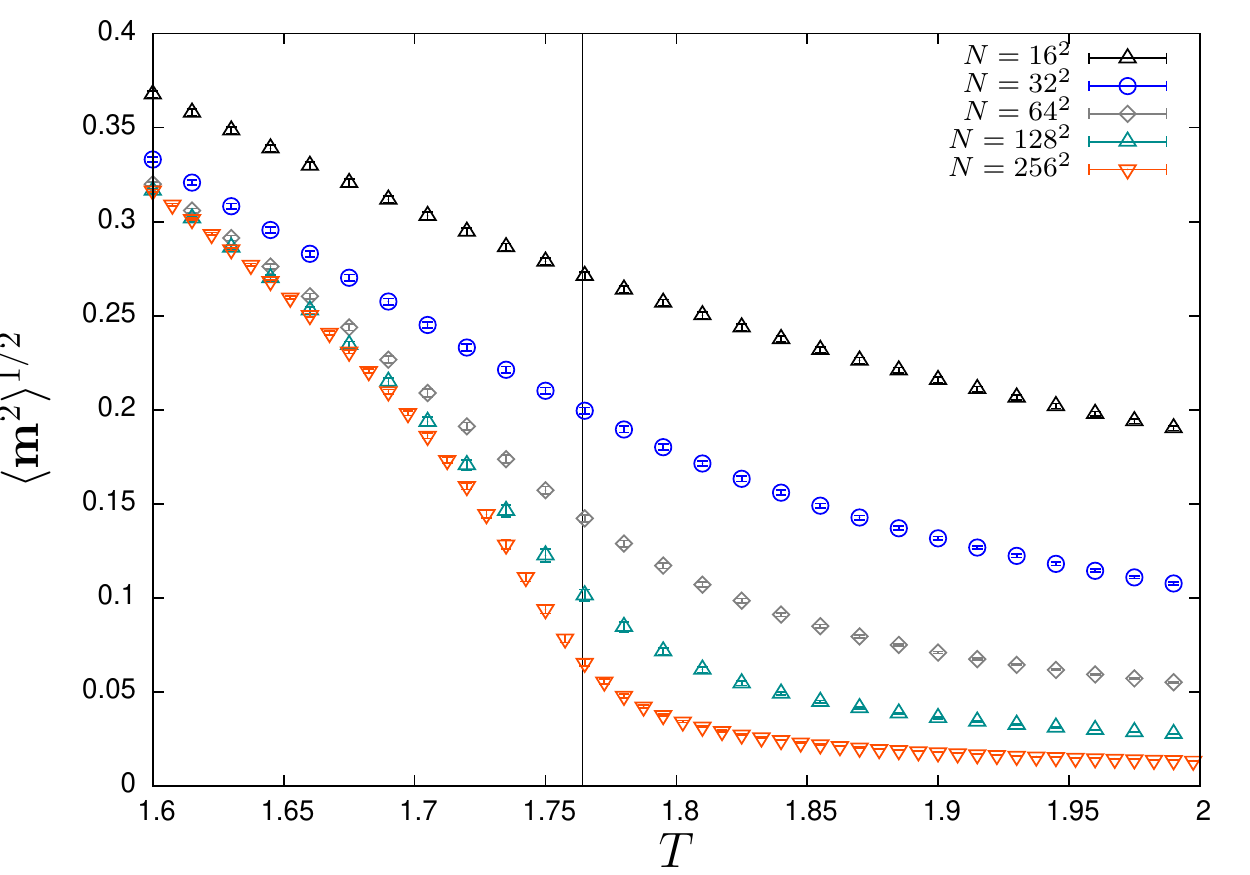} 
\caption{Square root of the average squared magnetization versus temperature for $\rho=10/3$ (Right), in the non-mean-field long-range regime. As $N$ increases, the low temperature phase exhibits finite spontaneous magnetization.}
\label{figrhonmfabsmag}
\end{center}   
\end{figure}

In Fig.   \ref{figrhonmfU4}, next to the main plot, we show in the insets  the $U_4$ rescaling at $\rho=10/3$ both for lattices with periodic and free boundary conditions. In the latter case  $T_c=1.53(2)$ and $\bar\nu=2.20(3)$,
the exponent being consistent with the FSS value from lattices with PBC. Also the susceptibility exponent turns out to be larger than its mean-field value, respectively, $\gamma=1.42(7)$ and $1.45(5)$. We, further, observe a low-$T$ magnetized state, $\<\mn m^2\>\ne 0$, almost insensitive to size, as in the mean-field case, cf. Fig. \ref{figrhonmfabsmag}.

{\it Comparison with 3D critical exponents.} In Figs. \ref{fig:nus}, \ref{fig:gammas} we also compare the values of the $\nu$ and $\gamma$ exponents at $\rho=3.307933$  (for which the equivalent short-range dimension is $D=3$ according to equations \ref{eq:Dsr_etasr_vs_drho})with their value in the 3D  XY model, obtained from a state-of-the-art numerical analysis.\cite{Campostrini2006Critical}

In particular, for $\rho=3.307933$, we find $\gamma=1.42(7)$, $\bar \nu=2.18(2)$, $\beta=0.39(2)$ to be compared with the values $\gamma_{\rm 3D}=1.3178(2)$, $\bar\nu_{\rm 3D}=3~\nu_{\rm 3D}=2.0151(3)$, $\beta_{\rm 3D}=0.3486(1)$.
Apparently, apart from $\gamma$ displaying the largest statistical uncertainty, these values do not satisfy  the quantitative relationships following the LR-SR equivalence conjecture, cf.  Eq. (\ref{eq:Dsr_etasr_vs_drho}) and Eq. (\ref{barnu_nurho}), even though their difference is rather small
(a few percent), as can be appreciated looking at
Figs. \ref{fig:nus} and \ref{fig:gammas}. 
The lack of accuracy in the determination of $\gamma$ comes from the fact that it is 
very sensitive to the value of the critical temperature used (see section \ref{FSS_multi}).

 The comparison is not any better  choosing
$\rho=10/3$ corresponding to a spectral, rather than Euclidean, dimension $\bar d=3$.

\subsection{Short-range regime}
\label{SRRegime}

The cases $\rho=3.75, 3.875$ and $4.667$  have bee simulated and analyzed finding evidence that they belong to the KT universality class
of the 2D short-range XY model.
 They also display some peculiar features that we compare to the numerical fingerprints of the KT transition reported in subsection \ref{benchmark}.

{\it (1)}  {\em Scale invariance at the critical point.}
 $\qquad$ First of all we estimate the critical point. We can do it by FSS of the crossing points of $U_4(T,N)$. Such estimate is, however,
  more and more difficult as $\rho$ increases because
 the low $T$ behavior of  $U_4(T,N)$ is less and less size dependent than the high $T$ behavior as $N$ increases, as shown
 in Fig.  \ref{figrhosrU4}. 
 This appears to be  a 
 precursor  of the low $T$ size independence occurring in 2D at the KT transition, as we already mentioned. Nevertheless we obtain
 $T_c=1.62(2), 1.57(1)$ and $1.38(2)$ for $\rho=3.75,3.875$ and $4.667$ respectively.
  We can, otherwise, estimate $T_c$ by means of the fixed $U_4$ FSS method, yielding $T_c=1.63(1), 1,58(1)$ and $1.36(1)$, respectively. 
  One can notice that the estimates for $\rho=4.667$ do not coincide because of the mentioned limits of the  crossings method.

 {\it (2)} {\em Susceptibility scaling at criticality.} $\qquad$ In the 2D XY model the susceptibility at criticality
behaves like Eq. (\ref{chi_SR})
with $\eta(T_c)=1/4$.\cite{Amit1999Renormalisation}
We numerically interpolated Eq. (\ref{chi_SR}) at different temperatures for different $\rho$ in the candidate short-range regime and also for $\rho=10/3<\rho_{\rm sr}$. The behavior of $\eta(T)$ is reported in Fig. \ref{eta_T_SR} for all these cases. 
As a reference, in the figure we also display the critical temperature intervals (vertical stripes), as estimated by the FSS method at fixed $U_4$.
 For all $\rho\geq 3.75$ we find that the temperature at which $\eta=1/4$ is compatible, in the statistical error, 
 with our estimate for $T_c$. This is not the case, instead, for $\rho=10/3$. 
 This hints that the conjectured $\rho_{\rm sr}=3.75$ is actually the 
threshold between LR and SR universality classes. In terms of the SR-LR equivalence, formulated in Eq. (\ref{eq:Dsr_etasr_vs_drho}),
 this confirms that $\rho_{\rm sr}$ is 
equivalent to $D=2$. 

\begin{figure}[t!]                        
\begin{center} 
 \includegraphics[width=.49\textwidth]{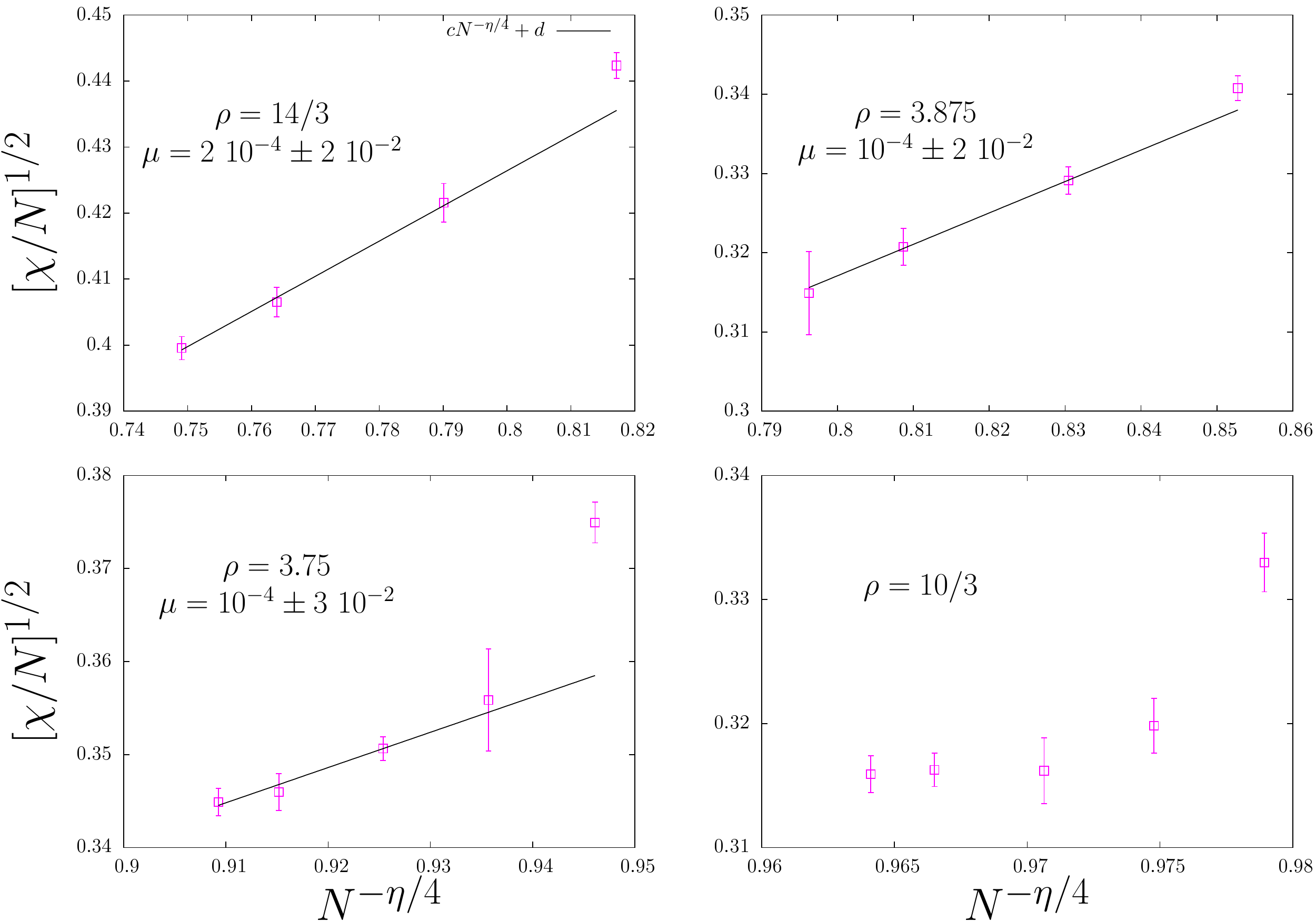} 
\caption{Scaling of $\sqrt{\chi/N}$ vs. $N^{-\eta/4}$ at $T<T_c$ for three values of $\rho$ in the SR regime and for $\rho=10/3<\rho_{\rm sr}$. The black lines are linear fits and the y-axis intercept $\mu$ is reported. For the three cases in the SR regime, the fitted value of $\mu$ is compatible with zero.}
\label{check_chi}
\end{center}   
\end{figure}

As a further confirmation of the fact that $\rho_{\rm sr}=3.75$, we present in Fig. \ref{check_chi} the behavior of the quantity $[\chi(T,N)/N]^{1/2}$ vs. $N^{-\eta(T)/4}$ for four values of $\rho$ and for $T$ values below the critical temperature. This allows for a self-consistency test of the scaling $\chi\sim N^{1-\eta/2}$ supposed in Fig. \ref{eta_T_SR}. In the SR regime, this quantity should behave as $[\chi/N]^{1/2}\sim N^{-\eta/4}$ for large $N$. As can be seen in Fig. \ref{check_chi}, this is verified for $\rho\geq 3.75$ and clearly not for $\rho=10/3$.

Once we are convinced that  for $\rho\geq 3.75$ the system is in the KT universality class, we can, vice-versa, interpolate a 
value of $T_c$ as the temperature at which
$\eta(T_c)=1/4$, yielding by a simple linear interpolation, $T_c=1.60(2), 1.56(1)$  and $1.37(1)$ for $\rho=3.75,3.875$ and $4.667$, respectively.

\begin{figure}[t!]                        
\begin{center} 
 \includegraphics[width=.49\textwidth]{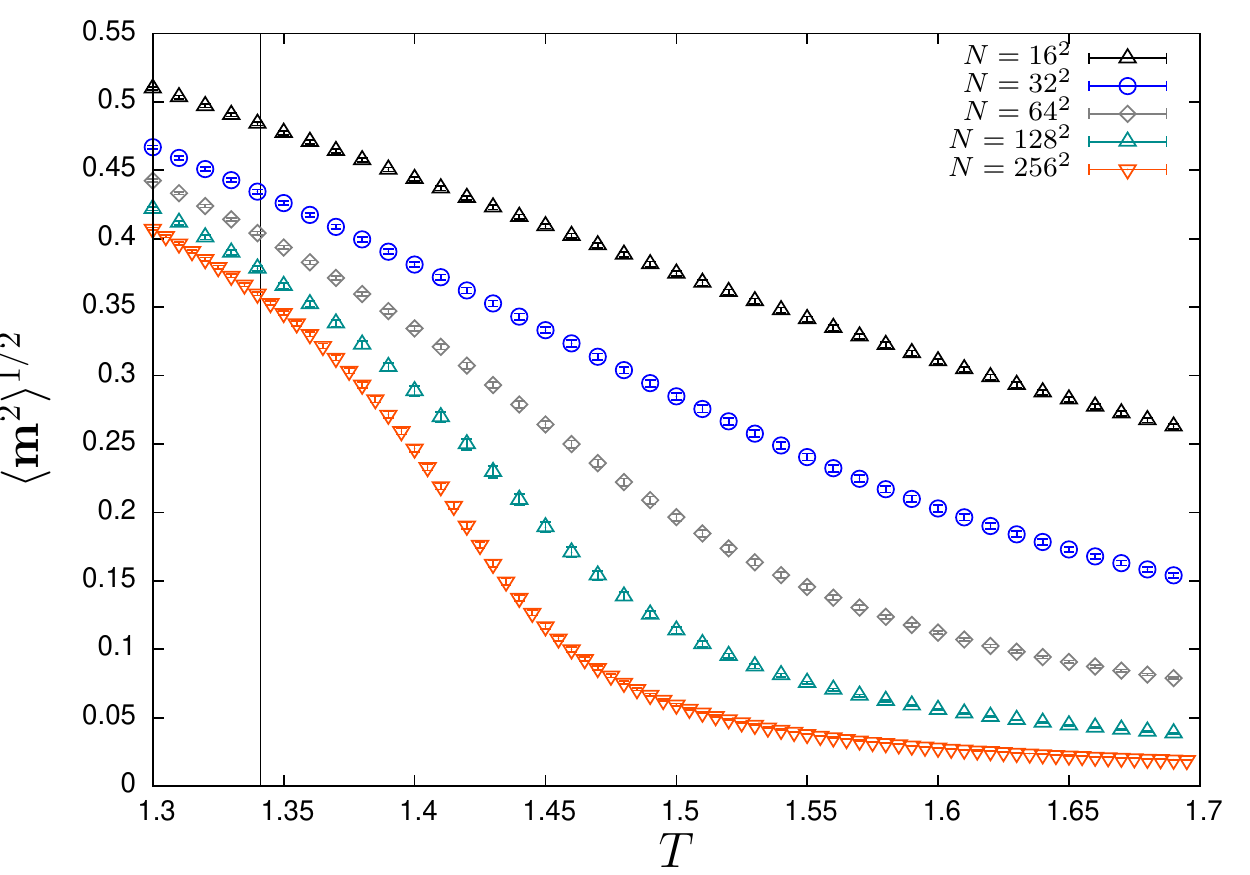} 
\caption{Magnetization versus temperature for $\rho=4.667$ in the short-range regime. In the low-$T$ phase the magnetization monotonously decreases with $N$.}
\label{figrhosrabsmag}
\end{center}   
\end{figure}

\begin{figure}[t!]                        
\begin{center} 
 \includegraphics[width=.49\textwidth]{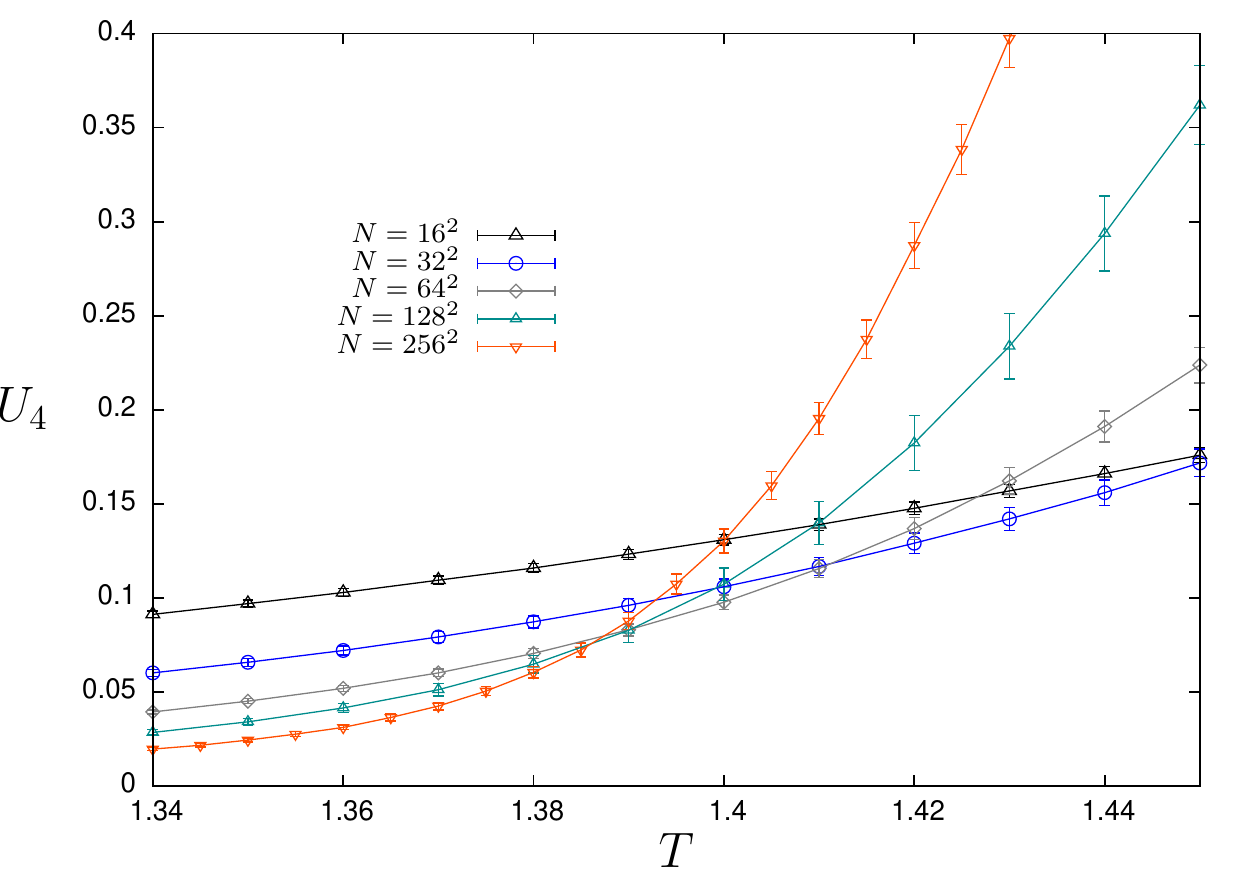} 
\caption{Binder cumulant versus temperature for $\rho=4.667$ in the short-range regime. There is evidence for the scale-invariance of this quantity for $T<T_c$ in the large-$N$ limit.}
\label{figrhosrU4}
\end{center}   
\end{figure}

 {\it (3)} {\em Magnetization in the low $T$ phase.}$\qquad$
The behavior of the magnetization below criticality is peculiar and might not be the same for all values of $\rho$ in the SR regime. 
Analytic results for XY spins on a random graph of spectral  dimension $\bar d$, \cite{Cassi1992Phase} indeed, prove that for $\bar d\leq 2$ the magnetization is zero in the thermodynamic limit ad it is non-zero for $\bar d>2$.
For $\rho$ large enough the squared magnetization goes to zero with the same  scaling of the susceptibility ($\sim N^{-1/16}$), see  Sec. \ref{benchmark} , Eq. (\ref{eq:m2_FSS_2D}) . For $\rho=4.667$, e.g., for which $\bar d=2$, we plot 
$({\overline{\langle \mn m^2
\rangle}})^{1/2}$ in Fig. \ref{figrhosrabsmag}.
As $\rho$ decreases below $4$ we  have  $\bar d >2$.
This implies a non-zero asymptotic value for  ${\overline{\langle \mn m \rangle^2}}$ as $\rho<4$ and, thus, different scalings for
$\chi$ and $({\overline{\langle  \mn m^2 \rangle}})^{1/2}$.
  In particular, the L\'evy graph with $\rho=3.875$ has
 $\bar d=4/(\rho-2)=2.1333\ldots$ and $\rho=3.75$ has spectral dimension $\bar d=2.2857\ldots$.  
 Unfortunately, because of the very slow scaling of the susceptibility it is rather hard to tell whether the asymptotic limit of the magnetization is compatible with a strictly positive value.
As $\rho$ is near the SR threshold $\rho=3.75$ we can, actually, not detect any relevant discrepancy with the $\chi$ scalings reported in Fig. \ref{check_chi}.

{\it (4)} The specific heat is not divergent nor discontinuous at the transition, cf., Fig. \ref{figcV}, right panel. 

{\it (5)} {\em Kosterlitz-Thouless law} $\qquad$ We have estimated the critical temperature in the SR regime using Eq. (\ref{eq:KTscaling}) at different sizes and taking the FSS of the fit parameter estimates, as we did for the square lattice case in subsection \ref{benchmark}. We obtain $T_c=1.59(1)$ for $\rho=3.875$ and
$T_c=1.34(2)$ for $\rho=4.667$. These estimates agree with the ones obtained from the FSS of the temperature at which the system presents a given value of $U_4$, $T_f(U_4,N)$. 

%


We summarize our results in all regimes  in Tab. \ref{criticalexponents}.



\section{Conclusions and perspectives}
\label{Conclusions}

The outcome of 
extensive numerical simulations on the 2D L\'evy lattice yield evidence 
for three different critical regimes, corresponding to given intervals in the L\'evy exponent $\rho$ governing the topology of the graphs: short-range $\rho\in[\rho_{\rm sr},\infty)$, non-mean-field long-range $\rho\in(\rho_{\rm mf},\rho_{\rm sr})$ and mean-field $\rho\in[0,\rho_{\rm mf}]$, 
with $\rho_{\rm mf}=3$ and $\rho_{\rm sr}=3.75$.
The SR threshold value has been determined in Sec. \ref{SRRegime}, where 
 we found  evidence that $\eta=1/4$ for
 $\rho\geq 3.75$ from the susceptibility scaling.

Studying the spectral dimension we verified that its expression, Eq. (\ref{eqbard}),  holding for fully connected long-range models still holds in the dilute case. The identification $\bar d(\rho)=2d/(\rho-d)$  appears, indeed, to be confirmed by a numerical estimation, see figure \ref{figdrho} and App. \ref{App:A}. 
 Furthermore, for $\rho\to d$, $\bar d$ diverges. An infinite spectral dimension, indeed, occurs in the Bethe lattice limit and, generally, 
 in any graphs not satisfying the polynomial growth condition. \cite{Burioni2005Random} 
The spectral dimension does not depend on the symmetry of the system but only on the topology of the graph. 

In the mean-field regime we measured the critical exponents that we found always consistent with the mean-field values
$\gamma=1$, $\alpha=0$ and $\bar\nu=2$. The latter is the correlation volume exponent, related with the correlation length exponent that we found always consistent with their mean-field values $\nu=\bar\nu/D_{\rm u}=1/2$. These exponents agree with the already mentioned theoretical predictions for the $D$-dimensional equivalent model in the mean-field regime: $\nu_\rho$, $\eta_\rho$, $\gamma_\rho$  
(see subsection \ref{meanfieldregime}). In the long-range non-mean field regime, instead, we find a continuous phase transition with different critical exponents and a low-temperature phase exhibiting spontaneous symmetry breaking. Finally, we report evidences for the onset of a KT-like transition  in the short-range regime for $\rho\geq \rho_{\rm sr}=2+d-\eta_{\rm sr}(2)=3.75$.  
In this regime we have the value $\rho=4$ corresponds to a spectral dimension $\bar d=2$. It is known \cite{Cassi1992Phase} that the XY model exhibits zero magnetization in graphs with $\bar d=2$ (i.e., for $\rho\ge 4$, see Eq. (\ref{eqbard})),  whereas for $\bar d>2$ ($\rho<4$) a finite magnetization should occur. \cite{Burioni1999Inverse} 
 Due to the very slow FSS of the magnetization ($\sim N^{-1/16}$ if the asymptotic value is zero), however, and because of the fact that for
  $\rho\lesssim4$ the asymptotic magnetization is expected to be small,  we have not been able to identify  a spontaneous
 $O(2)$ symmetry breaking  for $\rho<4$ with the simulated sizes.

For each value of $\rho$, the critical behavior can be conjectured to be in a one-to-one correspondence
with a short-range XY model in $D$ dimensions. This short-range effective dimension is exactly $D=d=2$ for  $\rho\geq\rho_{\rm sr}(d)$,
 and $D=2d/(\rho-d)$ for $\rho\in(d,3/2d)$, cf. Eq. (\ref{eq:Dsr_vs_drho}), in the mean-field regime. As  $\rho\to d$, $D$ tends to infinity. This is the value of $\rho$ for which the  fully connected version of the model  displays a divergent energy. 

The most delicate regime is the non-mean-field long-range one, for which  Eq. (\ref{eq:Dsr_vs_drho}) does not hold anymore and 
 the dimensional relationship derived from the SR-LR equivalence hypothesis is conjectured to be given by Eq. (\ref{eq:Dsr_etasr_vs_drho}).
 This can be derived, e.g.,  from a free energy scaling argument\cite{Banos2012Correspondence} or by requiring the exact match with the SR regime
 at $\rho_{\rm sr}$.\cite{Leuzzi2013}
So far  Eq. (\ref{eq:Dsr_etasr_vs_drho}) has been carefully tested in 1D
 L\'evy Ising spin-glasses for $\rho>\rho_{\rm mf}(1)=4/3$, verifying
 that the equivalent short-range critical behaviors are actually
 consistent both for $D=3$ ($\rho=1.792$) and for $D=4$ ($\rho=1.58$),
 but the compatibility is better the higher
 $D$ ($D_u=6$ in the spin-glass case).\cite{Banos2012Correspondence} 
 In the 2D (fully connected) ordered
 Ising model at $\rho=1.6546$ and $1.875$ that,  
 according to Eq. (\ref{eq:Dsr_etasr_vs_drho}), should correspond,
 respectively, 2D and 3D numerical estimates of
 critical exponents are consistent nearer to the mean-field threshold
 (in 3D) but for $\rho=1.875$ they do not appear compatible anymore with
 the 2D model.   \cite{Angelini12}
 
In the 2D XY model, we do not observe a strong disagreement for $D(\rho)=3$, that is for $\rho=3.307933$
but our more refined estimates are not compatible with the 3D results within the statistical error. Indeed, we obtain
 $\gamma=1.42(7), \bar \nu=2.18(2)$ and $\beta=0.39(2)$ to be compared with the values $\gamma_{\rm 3D}=1.3178(2)$, $\bar\nu_{\rm 3D}=2.0151(3)$
 and $\beta_{\rm 3D}=0.3486(1)$ of Ref. [\onlinecite{Campostrini2006Critical}].
We stress that the 2D limit is quite peculiar due to the uncommon specific critical properties of the KT transition, where the  low temperature  phase is unmagnetized
 and it is critical   at all $T\leq T_c$ with temperature dependent critical exponents and where $\chi$ and $\xi$ diverge exponentially
rather than power law and the very definition of $\bar \nu=\nu/2$ and $\gamma$ lacks, for $\rho<\rho_{\rm sr} =3.75$.
The ``3D'' values nevertheless seem to be not too far away from the approximated SR-LR correspondence 
expressed by Eqs. (\ref{eq:Dsr_etasr_vs_drho}), (\ref{barnu_nurho}) and (\ref{gamma_barnu}).

We have further determined that the spectral dimension is related with the dimension $D$ of a short-range lattice equivalent to the L\'evy lattice for what concerns the critical behavior. The two are identical in the mean-field regime, $\rho\leq 3$, cf. Eq. (\ref{eq:Dsr_vs_drho}) and Eq. (\ref{eqbard}). In this regime the structure of the graph alone is enough to determine the universality class of the system, independently of the symmetries of the system variables. 
Beyond the mean-field threshold symmetries of the specific system defined on the graph become relevant and
the identification does not hold anymore until $\rho\geq 4$ and the graph is by all means a bi-dimensional lattice: $\bar d=d=2$.  
The  result $D=\bar d$, valid in the mean-filed regime, implies that the critical behavior of the $O(2)$ model on a graph characterized by a spectral dimension $\bar d$ coincides with the one of the short-range  $\bar d$-dimensional $O(2)$ model, allowing for a deeper understanding of the physics of interacting systems on non-regular structures and extending the known universal properties of the spectral dimension of the $O(n\to\infty)$ model \cite{Burioni2000ntoinfty} to finite $n$.

Summarizing, we have found that the critical properties of the $O(2)$ model in a graph can be divided in three regimes characterized by the spectral dimension of the graph. In the mean-field regime it
 plays the role of the dimension of a short-range model with common critical properties. In the infrared divergent long-range regime $\bar d$ and $D$ do not coincide but are, somewhat, related, though the conjectured relationship Eq. (\ref{eq:Dsr_vs_drho}) does not seem to hold for 
 $\rho\in  [\rho_{\rm mf}:\rho_{\rm sr}]$.
  As a perspective, we propose to investigate how the introduction of disorder and a different short-range kind of criticality would change this scenario. \cite{Burioni1997Geometrical} This is the object of an ongoing research.

\section*{Acknowledgments}
We acknowledge Raffaella Burioni, Martin Hasenbusch and Flaviano Moroni for very interesting discussions. 
We thank the QUonG initiative of the INFN APE group for the use of their GPU cluster.  
The research leading to these results has received funding from
 the Italian Ministry of Education, University
and Research under the Basic Research Investigation
Fund (FIRB/2008) program/CINECA grant code
RBFR08M3P4 and under the PRIN2010 program, grant code 2010HXAW77-008 and 
from  the People Programme (Marie Curie Actions) of the European Union's Seventh Framework Programme FP7/2007-2013/ under REA grant agreement n¡ 290038, NETADIS project.

\appendix
\section{Spectral density estimation}
\label{App:A}

In order to estimate $\bar d$ we have studied the histogram of return times of a collectivity of $4 \times 10^4$ random walkers averaging over $20$ different finite size realizations of 2D dilute graphs with power $\rho=\infty$, $5$, $4.5$, $4$, $3.8$, $11/3$ and $10/3$. We have performed random walks on lattices of size $N=128^2$, $196^2$, $512^2$ and, for the smallest value of $\rho$, $N=768^2$. We have taken average coordination numbers $z=4$ and, for a comparison, $z=8$. Changing the average connectivity of the sites do not change the spectral dimension, within the statistical errors, at any given size.
Finite size effects are there, instead, as $\rho$ decreases, as discussed in Sec. \ref{spectraldimension}.

\begin{figure}[t!]           
\begin{center} 
 \includegraphics[height=6cm]{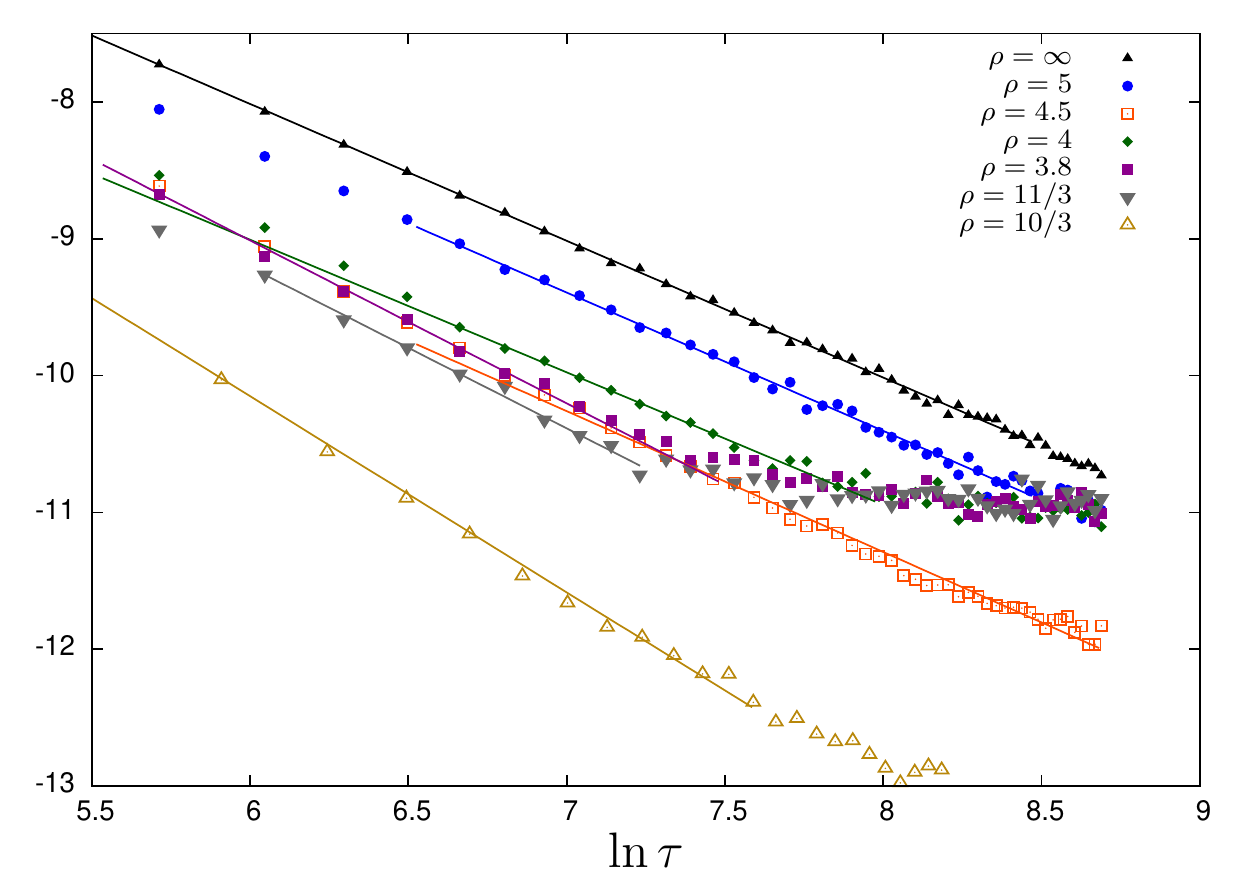} 
\caption{Logarithm of the histogram of self-return times of a random walker in dilute 2D graphs of size $N=196^2$ with different values of $\rho$. Points are the numerical measures, while lines are the fits with a linear function (their endpoints indicate the corresponding fit intervals). The slope of the fits is $-1/2$ times the spectral dimension (see figure \ref{figdrho}).}
\label{figdfits}
\end{center}   
\end{figure}

The histogram of return times is proportional to  the probability of self-return of a random walker in the graph after a time $\tau$:  $P(\tau)$. For large values of $\rho>2+d$ and in the square lattice case the estimation of $\bar d$ is very accurate since the function $P$ presents a very clear power-law behavior even at large times and finite-size effects are not an issue. For smaller values of $\rho$, however, such a measure becomes less and less accurate. This is due to the presence of  ``shortcuts'' on the graph that take the walker to the boundaries of the original lattice, where the probability $P$ is overestimated, and to the existence of low-connected nodes. To cope with these drawbacks, for $\rho>4$ our random walkers start from the center of the original finite-size 2D lattice, while for $\rho\le 4$, each realization of the walker starts from a node which is chosen at random between the subset of nodes whose degree is larger than two. In Fig. \ref{figdfits} we present the $P$ histograms (up to an arbitrary, $\rho$-dependent constant) for each studied value of $\rho$, from which the data of Fig. \ref{figdrho} have been inferred.

\section{Details of the algorithm}
\label{App:B}

We present here a  detailed description of the algorithm used: a home-made high-performance parallel code for the Monte Carlo dynamics of spin models defined on general networks. The software  is developed for GPUs architecture and it has been  developed with the CUDA programming model. 
A single-spin flip Metropolis update has been used, with non-connected spins being updated in parallel by different GPU cores.
Though this might not be the optimal algorithm for the specific case of the long-range ferromagnet, the kind of parallel programming we propose has rather competitive performances and, on top of that, is straightforwardly exportable to any kind of system with continuous variables, including models with random bonds and fields.

\subsubsection{Graph coloring}
This procedure requires the {\it coloring} of  each realization of the randomly generated graph  before dynamics starts. 
The graph nodes are colored with the same color if they are not connected to each other.
During the simulation, sites with a common color are Metropolis-updated synchronously, and subsets of the set of vertexes corresponding to different colors are processed sequentially on each MCS. This is a generalization of the so called Red-Black Gauss Seidel algorithm used in the parallelization of spin operations in bipartite graphs, as hyper-cubic lattices. 

\begin{figure}[t!]                        
\begin{center} 
 \includegraphics[width=.99\columnwidth]{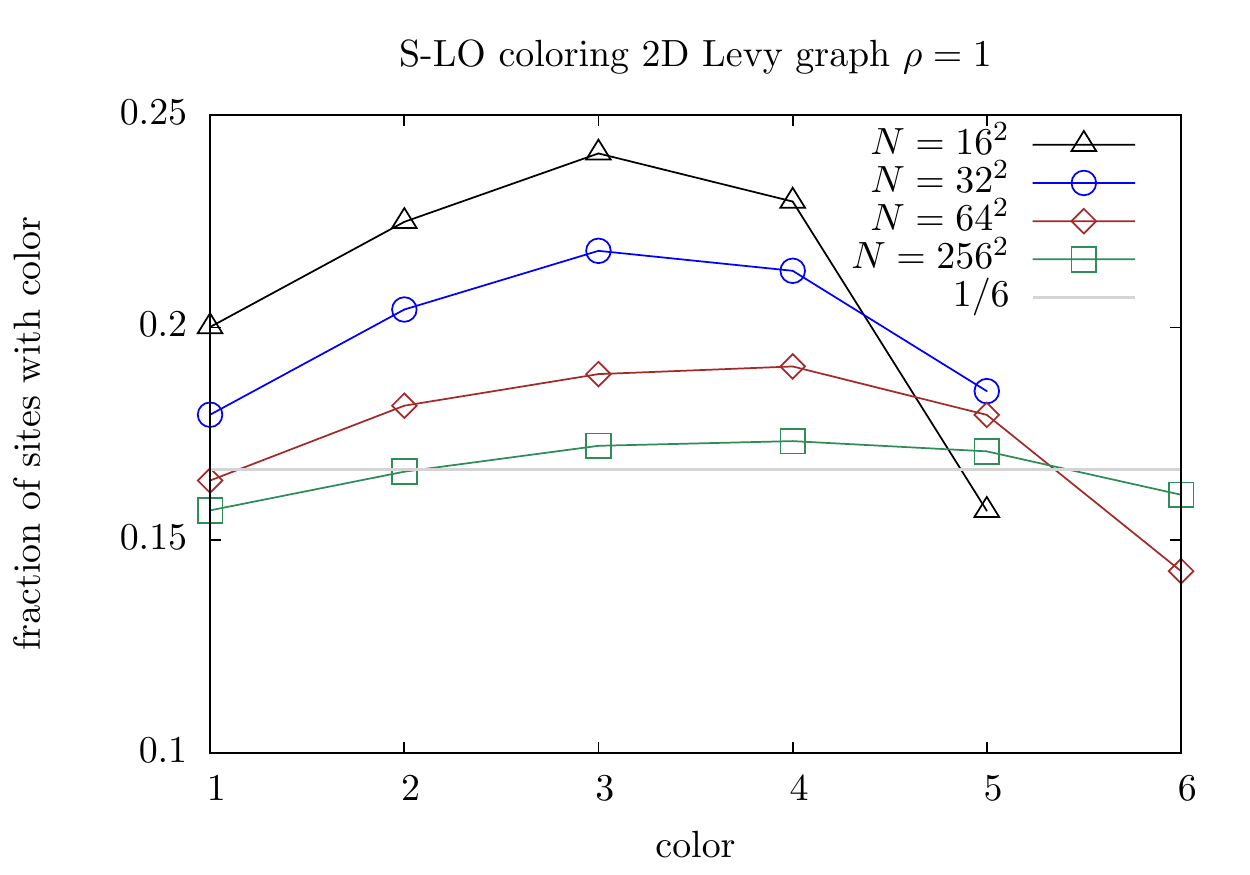} 
\caption{Coloring a dilute L\'evy 2D graph with power $\rho=1$ and $N=L^2=2^{8-14}$ nodes with the SLO algorithm. The larger the graph, the more homogeneous is the partition of the set of graph nodes in subsets corresponding to different colors.}
\label{figcoloring}
\end{center}   
\end{figure}

We approximately color the graph using a variant of the Smallest-Last-Ordering (SLO) algorithm, \cite{Matula1983Smallestlast,BerganzaUntitled} costing $O(N)$.
 For the simulated sizes  ($N\le 2^{16}$) the number of colors (equal to two in the $\rho=\infty$ case) turns
out to be never larger than $6$. 
As one can see in  Fig. \ref{figcoloring}, with our coloring procedure the distribution of non-interacting sets 
 becomes
more and more homogeneous as $N$ increases, thus automatically enhancing  the algorithm efficiency.\cite{BerganzaUntitled}

\subsubsection{Improved equilibrium dynamics}
In our code, besides the Metropolis algorithm, also the Parallel Tempering (PT) \cite{Hukushima1996Exchange} and Over-relaxation (OR) \cite{Creutz1987Overrelaxation} algorithms  are implemented. Both algorithms reduce the correlation time of the Monte-Carlo Markov processes and improve the equilibration. \footnote{An alternative optimization can be realized by adopting cluster accelerating algorithms  adapted to dilute long-range systems, see e.g., 
\cite{Wolff, Luijten1995, Luijten1996}
We do not implement such algorithm in the present work.}
 
\begin{figure}[t!]                        
\begin{center} 
 \includegraphics[width=.99\columnwidth]{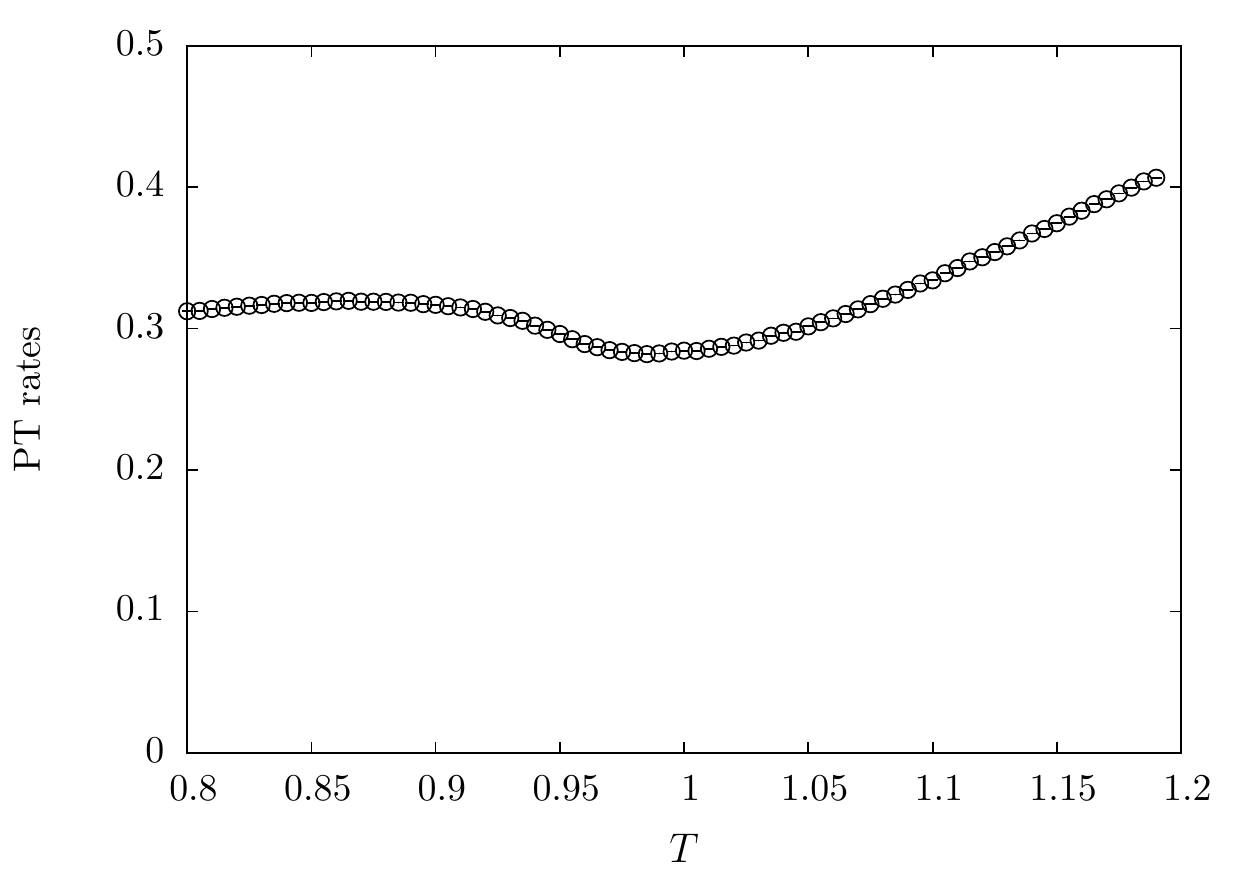} 
\caption{Exchange rate between nearby heat baths in the PT algorithm in a square lattice with $N=2^{16}$ sites. The distance between consecutive temperatures is $\Delta T=0.005$.}
\label{figPTrates}
\end{center}   
\end{figure}

PT swap attempts are performed (in CPU) every MCS, with replicas at different temperatures being updated in parallel, as explained in Ref. \onlinecite{Weigel2012Performance}. Fig. \ref{figPTrates} illustrates the rate of PT swaps between configurations with adjacent temperatures at fixed intervals of $\Delta T=0.005$, as a function of the temperature for a system with $N=2^{16}$ in the 2D square lattice, for which the critical temperature is known to be $Tc=0.8929...$ .

\subsubsection{Memory management}
We have used a storage of the degrees of freedom  in the {\it global device memory} of the GPU architecture, \cite{Weigel2012Performance,Ferrero2012qstate} each {\it thread} accessing the $O(2)$ angle of its corresponding graph node in such a way that sites with a common color are consecutive in the array, favoring {\it coalesced} memory access. Each {\it thread}, then, accesses  an array  in global memory, from which it reads the list of sites connected to the corresponding site. 
An independent  random number generator of the Fibonacci type \cite{Weigel2012Performance}
is associated to each {\it device thread}. 
We used double floating-point precision for storing observables, and single precision for the calculation of the trigonometric functions in the evaluation of the energy and magnetization of each site. In the latter case we adopted the special {\sf fast$\_$math} function of the GPU architecture, a faster routine specific of the GPU architecture. 

\begin{table}[t!]
\begin{center}
\begin{small}
\begin{tabular}{ccc|c}
algorithm & precision & { trig. function}  &  time per spin [ns]\\
\hline
MET + PT & single & {\sf fast} & 1.88 \\
MET & single & {\sf fast} & 0.635 \\
MET & single & {\sf cosf} & 0.865 \\
OR  & single & {\sf fast} & 0.36 \\
OR  & single & {\sf cosf} & 0.54 \\
\hline
\end{tabular}
\caption{Computational time of different algorithms per spin. It is the total computation time of a run divided by $N$, by the number of copies $N_T$ at different temperatures  in the PT algorithm and by the number of MCS's.
}
\label{tabletimes}
\end{small}
\end{center}
\end{table}

\begin{figure}[t!]                        
\begin{center} 
 \includegraphics[width=.49\textwidth]{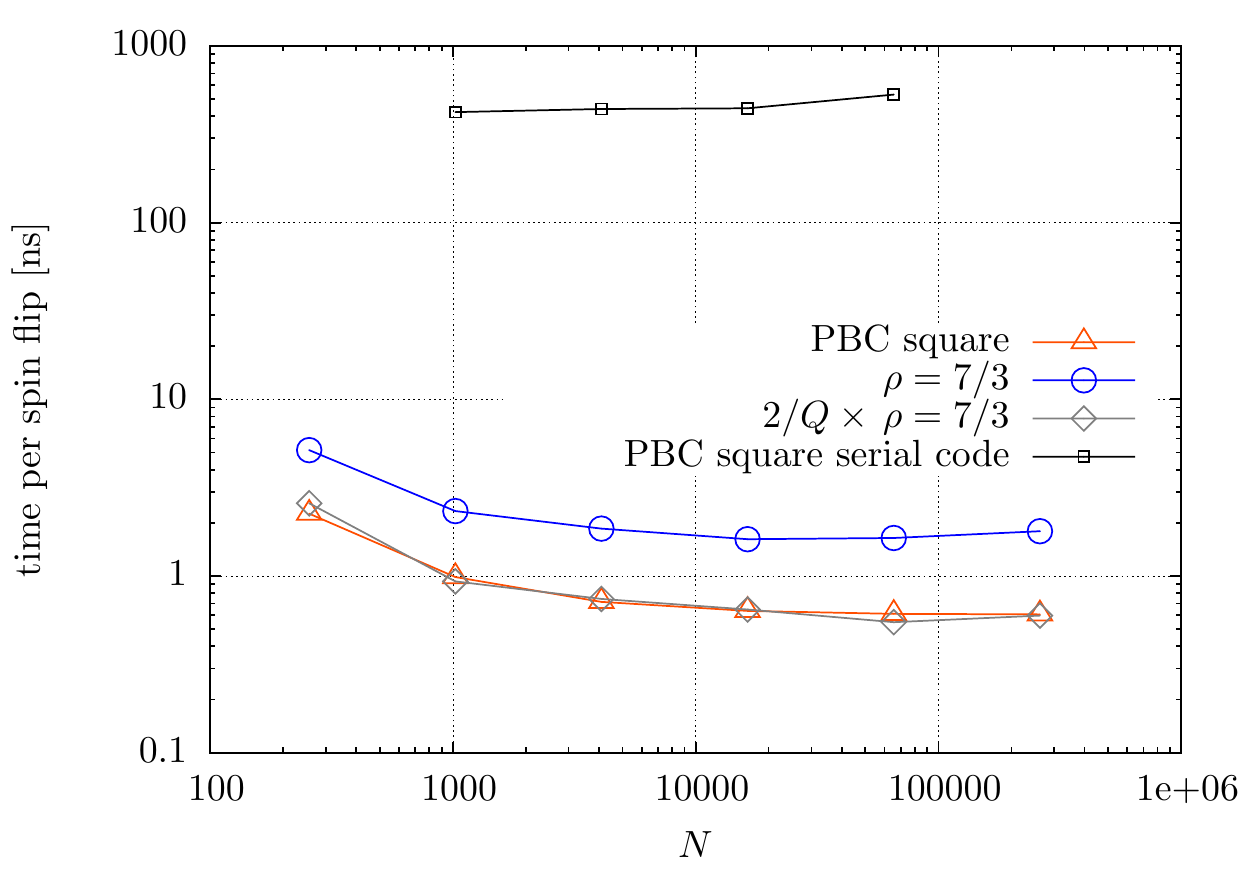} 
\caption{Computational time  per spin of the Metropolis algorithm versus $N$ for  the square nearest-neighbor lattice and $\rho=7/3$ L\'evy lattice. 
The serial-CPU run is shown for comparison: a speedup of several hundreds of times can be observed.}
\label{figtimes}
\end{center}
\end{figure}                          

\subsubsection{Computational speed}
We now present some details about the performance of our algorithm, referring to a calculation performed in an nVidia GPU GTX480 Fermi card. In Tab. \ref{tabletimes} the reader may find the computation time per spin involved in the Metropolis and OR algorithms in a square lattice with $N=2^{14}$ sites and PBC, for different choices of the floating point precision and of the routine used for the computation of trigonometric operations. In Fig. \ref{figtimes} a comparison of the computation time 
 for the PBC square and L{\'e}vy lattice with $\rho=7/3$ 
 with different sizes is shown. Since in a general graph colored with $Q$ colors our algorithm is nearly $Q/2$ times slower than the code in the square lattice, we also show $2/Q$ times the computation time in the L\'evy graph for comparison with the square lattice case. The minimum time peaks for Metropolis and OR algorithms are 0.55ns and 0.36ns respectively, a  mark which is competitive  with state-of-the-art highly-optimized GPU simulations of spin-glasses, \cite{Yavorskii2012Optimized} (a direct comparison is not possible since their benchmark refers to the $O(3)$ model) and represents a speedup of several hundred times with respect to a serial C-code running on an Intel i7 CPU with 2.67GHz. 
All of the simulations  in this work have been performed using single precision (4 bytes) for the storing of floating-point numbers and the {\sf fast\_math} CUDA functions.  We ran simulations changing both precision and trigonometric routines, and introducing OR sweeps, without finding essential accuracy improvements. An upgraded and generalized version of this algorithm, designed for the study of random laser modes \cite{Cao03,Wiersma08,Leuzzi09,Conti2011Complexity} in arbitrary topologies, will be extensively reviewed and presented in a forthcoming work. \cite{BerganzaUntitled}

\section{Finite size scaling analysis of critical parameters}
\label{FSS_multi}
		\begin{center}{\em Critical temperature.}\end{center}

For each value of $\rho$, we have estimated $T_c$  both from the FSS of the crossing points of $U_4$ at different sizes (see, e.g, Figs.  \ref{figrhomf1U4}, \ref{figrhonmfU4}) and from the FSS of  the temperature at which a finite size system exhibits a given value $U_4$ of the Binder cumulant: 
\beq
T_{f}(U_4,N) = T_c + A\,N^{-x}
\label{eq:Tfsscaling}
\eeq
being $T_c=T_f(U_4,\infty)$ the critical temperature, independent from the specific $U_4$ value chosen for the fit, and $x$, a quantity in principle depending on $U_4$, and that can be identified with $1/\bar\nu$ for $\rho<3.75$ (cf. Fig. \ref{fig:Ts}). In order to find $T_c$: (i) we take different values of the Binder cumulant $U_4^{(j)}$, $j=1\ldots n_d$ in a reasonable range around the critical region, (ii) we construct $n_d$ apart  data sets $\{T_f(U_4^{(j)},N)\}_j$ and (iii) we interpolate 
all datasets simultaneously with  common parameter $T_c$ and set-depending parameters $A(j)$, $x(j)$. The resulting temperatures are reported in  Tab. \ref{criticalexponents} and plotted in Fig. \ref{fig:Ts}  together with the $T_c$ estimated from the FSS of the crossing points. In practice, for $\rho<\rho_{\rm sr}=3.75$, we fix $x$ to be common to all $U_4$ values, while for $\rho>3.75$ it is $U_4$-dependent. 

		\begin{center}{\em Correlation volume exponent}\end{center}

Besides estimating $\bar \nu$ from the interpolation with Eq. (\ref{eq:Tfsscaling}), in order to have a more precise determination 
we estimated the correlation volume exponent from the logarithm of  temperature derivative of the binder $\dot U_4(T,N)$ at fixed $U_4$.
Performing a simultaneous FSS fit over  apart datasets at different values of the Binder cumulant
with the law

\beq
\ln \dot U_4(T_j,N) = c_j + \frac{1}{\bar\nu}\,\ln N
\label{eq:scaling_lndotU4}
\eeq
and with a common value of $\bar \nu$ for all datasets. We obtain the results plotted in Fig. 
 \ref{fig:nus} and reported in Tab. \ref{criticalexponents}.

		\begin{center}{\em Susceptibility exponent}\end{center}

The $\gamma /\bar\nu$ exponent has been determined from the FSS (\ref{chi_scale}) in the approximated form:
\beq
\ln \chi(T,N) = {\rm cte}(T)+x(T)\ln N
\label{eq:chiscaling}
\eeq
where $x(T_c)$ can be identified with $\gamma/\bar\nu$ in the MF, LR regimes, and $x(T)$ can be identified with $1-\eta(T)/2$ in the SR regime, when $T\le T_c$. We have interpolated $x(T)$ for several values of the temperature in the scaling region, (as shown in figure \ref{eta_T_SR}), and estimated the values of $\gamma/\bar\nu$ from the values of $x(T)$ with $T$ in the error interval of $T_c$, estimated as explained above.   

 Finally, $\gamma$ is  obtained by multiplying the interpolated  $\gamma/\bar\nu$ by the $\bar\nu$ obtained from the $U_4$ fit. 
The resulting values of $\gamma$ and of $\gamma/\bar\nu$ so computed are shown in figure \ref{fig:gammas}. 


\begin{center}{\em Magnetization exponent $\beta$}\end{center}

The exponent $\beta$ is estimated in a similar way as done for the $\gamma$ in Eq. \ref{eq:chiscaling}, 
assuming the FSS:

\beq
\frac{1}{2}\ln {\overline{\< \mn m ^2\>}}(T,N) = {\rm cte}(T) - x(T)\ln N \qquad T < 0
\label{eq:mscaling}
\eeq
identifying $x(T_c)$ with $\beta/\bar\nu$  yields the  estimates  reported in Tab. \ref{criticalexponents}. It is interesting to remark that the (Rushbrooke-Widom) scaling relation between critical exponents $\gamma/\bar\nu=1-2\beta/\bar\nu$ is satisfied. The quantity  $1-2\beta/\bar\nu$ is reported in figure \ref{fig:gammas} for several values of $\rho$, illustrating the validity of the Rushbrooke-Widom  relation.

\bibliography{berganza}
\bibliographystyle{apsrev4-1}

\end{document}